\documentclass[preprint2]{aastex}

\newcommand{\asecx}{$^{\prime \prime} \!\!$}
\newcommand{\asec}{$^{\prime \prime}$ }
\newcommand{\degx}{$^{\circ} $ }
\newcommand{\degy}{$^{\circ}$}
\newcommand{\kms}{km~s$^{-1}$ }
\newcommand{\kmsx}{km~s$^{-1}$}
\newcommand{\Oiii}{[OIII] $\lambda $5007 }
\newcommand{\Oiiix}{[OIII] $\lambda $4959 }
\newcommand{\Oii}{[OII] $\lambda $3727 }
\newcommand{\Siia}{[SII] $\lambda $6717 }
\newcommand{\Siib}{[SII] $\lambda $6731 }
\newcommand{\Hb}{H$\beta$ }
\newcommand{\Ha}{H$\alpha$ }

\input psfig.sty
\input epsf.sty

\def\plotonex#1#2{\centering \leavevmode \epsfxsize=#2 \epsfbox{#1}}
\def\plotoney#1{\centering \leavevmode \epsfxsize=.45\columnwidth \epsfbox{#1}}
\def\plottwo#1#2{\centering \leavevmode
\epsfxsize=.45\columnwidth \epsfbox{#1} \hfil
\epsfxsize=.45\columnwidth \epsfbox{#2}}

\begin{document}

\title{STIS Longslit Spectroscopy Of The Narrow Line Region Of NGC 4151. 
I. Kinematics and Emission Line Ratios\altaffilmark{1}}

\author{C. H. Nelson\altaffilmark{2}, D. Weistrop\altaffilmark{2}, 
J. B. Hutchings\altaffilmark{3}, D. M. Crenshaw\altaffilmark{4}, 
T. R. Gull\altaffilmark{5}, M. E. Kaiser\altaffilmark{6}, 
S. B. Kraemer\altaffilmark{4}, D. Lindler\altaffilmark{5}}

\altaffiltext{1}{Based on observations with the NASA/ESA \it Hubble Space 
Telescope, \rm obtained at the Space Telescope Science Institute, which 
is operated by AURA Inc under NASA contract NAS5-26555}

\altaffiltext{2}{Physics Dept., University of Nevada, Las Vegas, Box 4002, 
4505 Maryland Pkwy., Las Vegas, NV 89154, cnelson@physics.unlv.edu, 
weistrop@nevada.edu}

\altaffiltext{3}
{Dominion Astrophysical Observatory, National Research Council of Canada, 
5071 W. Saanich Rd., Victoria B.C. V8X 4M6, Canada}

\altaffiltext{4}
{Catholic University of America, NASA/Goddard Space Flight Center,
Code 681, Greenbelt, MD 20771}

\altaffiltext{5}{NASA/Goddard Space Flight Center,
Code 681, Greenbelt, MD 20771}

\altaffiltext{6}{Department of Physics and Astronomy, Johns Hopkins University,
Baltimore, MD 21218}

\begin{abstract}

Longslit spectra of the Seyfert galaxy NGC 4151 from the UV to near
infrared have been obtained with STIS to study the kinematics and
physical conditions in the NLR.  The kinematics show evidence for
three components, a low velocity system in normal disk rotation, a
high velocity system in radial outflow at a few hundred \kms relative
to the systemic velocity and an additional high velocity system also
in outflow with velocities up to 1400~\kmsx, in agreement with results
from STIS slitless spectroscopy (Hutchings {\it et al.}, 1998, Kaiser
$et~al.$, 1999, Hutchings {\it et al.}, 1999) We have explored two
simple kinematic models and suggest that radial outflow in the form of
a wind is the most likely explanation. We also present evidence indicating 
that the wind may be decelerating with distance from the nucleus.

We find that the emission line ratios along our slits are all entirely
consistent with photoionization from the nuclear continuum source. A
decrease in the \Oiii / \Hb and \Oiii / \Oii ratios suggests that the
density decreases with distance from the nucleus. This trend is borne
out by the [SII] ratios as well.  We find no strong evidence for
interaction between the radio jet and the NLR gas in either the
kinematics or the emission line ratios in agreement with the results
of Kaiser {\it et al.} (1999) who find no spatial coincidence of NLR
clouds and knots in the radio jet. These results are in contrast to
other recent studies of nearby AGN which find evidence for significant
interaction between the radio source and the NLR gas.

\end{abstract}


\clearpage

\section{Introduction}

Since the launch of the Hubble Space Telescope (HST) many 
imaging studies of the Narrow Line Regions (NLR) of active galactic
nuclei (AGN) have been carried out. These studies have shown that the
emission line gas often has a complex morphology, frequently taking the
form of a bicone centered on the galaxy nucleus.  ({\it e.g.} NGC
4151, Evans et al., 1993, Boksenberg {\it et al.}  1995, NGC 1068,
Evans {\it et al.} 1991, see also the archival study by Schmitt \&
Kinney 1996). In the standard model for an AGN, a dense molecular
torus with a radius of a few parsecs surrounds the nucleus and
collimates the radiation field ({\it e.g.} Antonucci 1993).
According to the model, differences in the continuum and emission line
spectra, which form the basis for classification of Seyferts and
other types of AGN, can be explained largely by differences in the
orientation of the torus to our line of sight.  For example, in type 1
Seyfert galaxies our viewing angle is close to the symmetry axis of
the torus allowing a direct view of the Broad Line Region (BLR) and
the nuclear continuum source, while in type 2 Seyfert galaxies our
vantage point lies closer to the plane of the torus which then blocks
a direct view of the inner regions.

In many instances the NLR morphology and kinematics appear
closely linked to the radio structure, particularly in Seyfert
galaxies with linear or jet-like radio sources. In these objects the
line emitting gas is often found to be cospatial with the radio jets
and there is also kinematic evidence for physical interaction between
the jets and the NLR gas (Capetti {\it et al.} 1999, Whittle {\it et
al.}, 1988). The suggestion has been made that expansion of the radio
plasma into the host galaxy's interstellar medium produces fast shock
waves which emit a hard continuum and ultimately provide the dominant
source of ionizing photons (Taylor, Dyson, \& Axon, 1992, Sutherland,
Bicknell \& Dopita, 1993).

The degree to which photoionization by a nuclear continuum source or
by autoionizing shocks contributes to the overall energetics of the
NLR has been the subject of some debate.  In principal one can
distinguish between them spectroscopically by studying the spatially
resolved kinematics and the physical conditions of the gas as revealed
by the relative intensities of specific emission lines.  The Space
Telescope Imaging Spectrograph (STIS) is ideally suited for this type
of study. We have therefore undertaken a detailed investigation of the
kinematics and physical conditions across the NLR of NGC 4151, one of
the nearest Seyfert galaxies.

Evidence for outflow and photo-ionization cones in the NLR of
NGC 4151 was presented by Schulz (1988, 1990) based on ground-based
longslit spectroscopy. Peculiar flat-topped and double-peaked emission line
profiles were observed to the SE and NW between 2\asec \ and 6\asec \
from the nucleus and are most consistent with outflow models. Schulz (1990)
suggests that the outflow is driven either by a wind related to the
active nucleus or by an expanding radio plasmon.

The NLR kinematics in NGC 4151 have been studied in detail using
slitless spectroscopy from STIS (Hutchings {\it et al.}, 1998, Kaiser
{\it et al.}, 1999, and Hutchings {\it et al.}, 1999).  These
observations reveal three distinct kinematic components: one
consisting of low velocity clouds ($|V-V_{sys}| \sim 100$ \kms),
primarily in the outer NLR following the rotation of the host galaxy
disk, a second consisting of moderately high velocity clouds
($|V-V_{sys}| \ge 400$ \kms) most likely associated with radial
outflow within the biconical morphology and a third component of
fainter but much higher velocity clouds ($|V-V_{sys}| \sim 1400$ \kms)
which is also outflowing but not restricted to the biconical flow of
the intermediate velocity component. No evidence for higher
velocities in the vicinity of the radio knots was found suggesting
that the radio jet has minimal influence on the NLR kinematics.

A somewhat different conclusion was drawn by Winge {\it et al.} (1999)
primarily using longslit spectroscopy with HST's Faint Object
Camera. They claim evidence for strong interaction between the radio
jet and the NLR gas.  Furthermore, after subtracting the influence of
the radio jet and galaxy rotation on the kinematics, they suggest that
the residual motion is the rotation of a thin disk of gas on nearly
Keplerian orbits beyond 0\asecx.5 (60 pc using their linear scale)
around an enclosed mass of $\rm 10^9 M_{\odot}$. Interior to 60 pc the
velocities turn over suggesting that the mass is extended, and, if
their interpretation is correct, they are able to place upper limits
on the mass of a nuclear black hole of $\sim 5 \times 10^7 \rm
M_{\odot}$.

In this paper we present the initial results from our low resolution,
longslit spectroscopy. A second paper presents a detailed
photoionization model using the emission line ratios presented here
(Kraemer {\it et al.} 1999, Paper II). Section \ref{obs} presents the
observations and describes the data reduction procedures including
correction for scattered light from the Seyfert nucleus. Section
\ref{ana} describes the results of the kinematic and preliminary line
ratio analyses. In section \ref{disc} we discuss the results in terms
of different NLR models. We summarize our results and conclusions in
section \ref{conc}.

\section{Observations and Data Reduction}
\label{obs}

\begin{figure}
\vspace*{-1.6truein}
\hspace*{-0.75truein}
\psfig{file=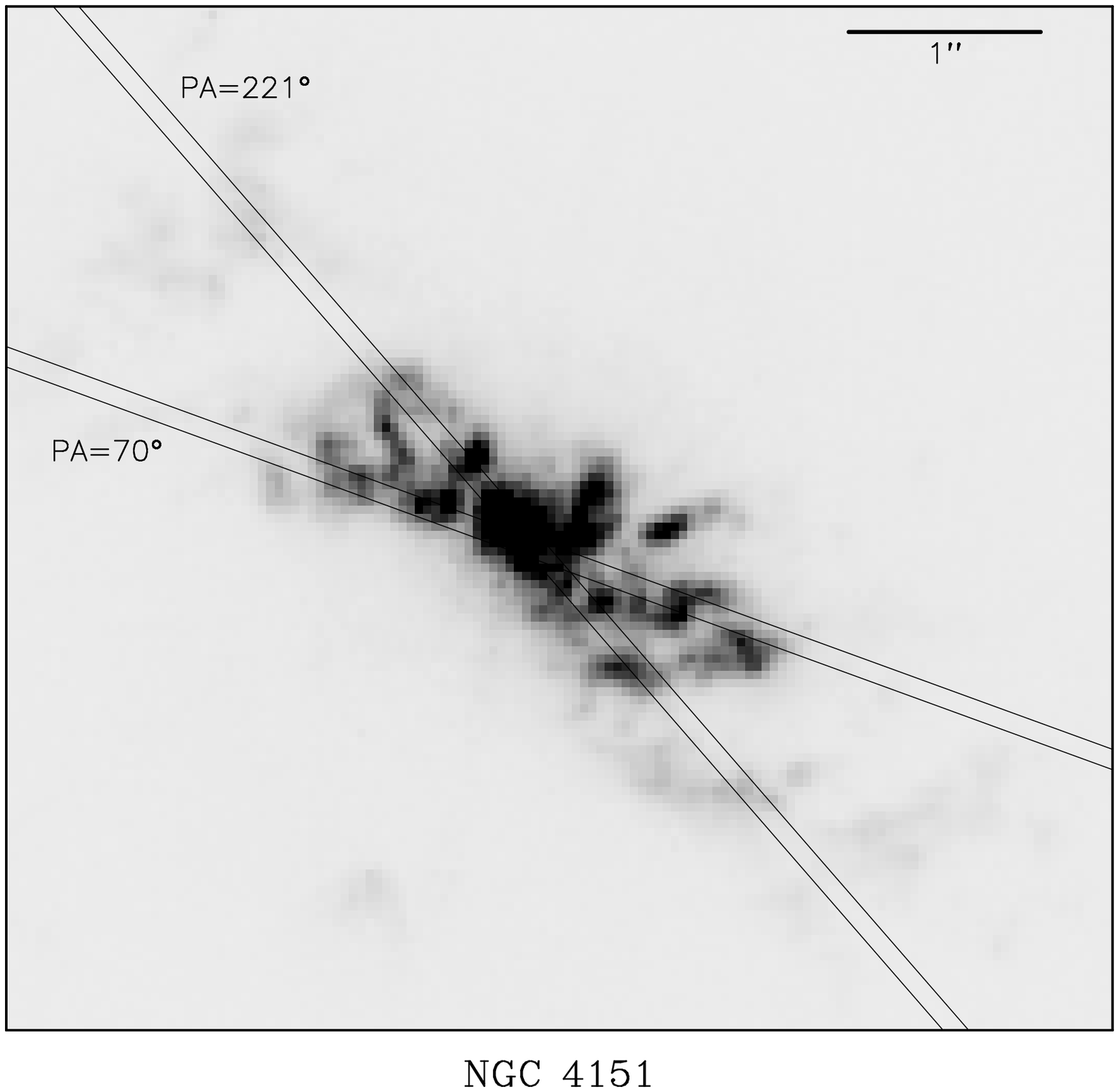}
\vspace{-1.0truein}
\figcaption{A continuum subtracted WFPC-2 F502N image of NGC 4151 is shown with 
the STIS longslit apertures superposed. North is up and East is at the left.
The slit at P.A. 221\degx goes through the nucleus while the one at P.A. 70\degx
passes 0\asecx.1 to the south.
\label{slits}}
\end{figure}

Longslit spectroscopy of NGC 4151 was obtained with STIS on board HST.
Four low dispersion gratings, G140L, G230LB, G430L and G750L, were
used producing spectra ranging from the UV at 1150 \AA \ to the
near-infrared at 10,270 \AA. Note that the G230LB mode, which uses the
CCD detector, was used instead of the G230L, due to the bright object
protection limits imposed on use of the MAMA detectors.  Two slit
alignments were chosen to cover regions of specific interest and as
many of the bright emission line clouds as possible. The first
position was chosen to pass through the nucleus at position angle
221\degy, while the second was offset from the nucleus by 0\asecx.1 to
the south at position angle 70\degy. Figure \ref{slits} shows the slit
apertures drawn on the WFPC-2 narrow band image of the \Oiii emission
line structure obtained from the HST archives (proposal ID 5124,
principal investigator H. Ford). The 0\asecx.1 slit was used to
preserve spectral resolution, given here for each of the four gratings
assuming an extended source (the emission line clouds are generally
resolved along the slit): 2.4 \AA \ for G140L, 2.7 \AA \ for G230LB,
5.5 \AA \ for G430L, and 9.8 \AA \ for G750L (Woodgate {\it et al.}
1998, Kimble {\it et al.} 1998, Baum {\it et al.} 1998). A log of the
observations is presented in Table 1.  One set of observations failed
and as a result no G140L spectrum was available for P.A. 70\degy.

The spectra were reduced using the IDL software developed at NASA's
Goddard Space Flight Center for the Instrument Definition Team
(Lindler {\it et al.} 1998). Cosmic ray hits were identified and
removed from observations using the CCD detector (G230LB, G430L, and
G750L) by combining the multiple images obtained at each visit in each
spectroscopic mode. Hot or warm pixels (identified in STIS dark
images) were replaced by interpolation in the dispersion
direction. Wavelength calibration exposures obtained after each
science observation were used to correct the wavelength scale for
zero-point shifts. The spectra were also geometrically rectified and
flux-calibrated to produce a constant wavelength along each column
(the spatial direction) and fluxes in units of ergs s$^{-1}$ cm$^{-2}$
\AA$^{-1}$ per cross-dispersion pixel. Spectra obtained at the same
position angle and spectroscopic mode were combined to increase the
signal-to-noise ratios.

The bright, unresolved Seyfert nucleus of NGC 4151 creates a number of
difficulties when trying to examine emission lines from the NLR close
in. Scattered light, largely from Airy rings imaged on the slit,
causes features of the nuclear point source spectrum to be
superimposed on fainter NLR features. These follow linear tracks
running nearly parallel to the dispersion, diverging slightly with
wavelength and can be detected as much as 20$-$30 pixels from the
nucleus (Bowers \& Baum, 1998). This is a particularly difficult
problem for measuring the Balmer lines in the NLR since the BLR lines
are strong and often have peculiar shapes which can influence the
continuum placement if not subtracted properly.  In addition, the
extended halo of the PSF must be modeled and subtracted. Furthermore, 
reflection of the bright nucleus in the CCD modes appears as a ghost
spectrum, which is displaced from the nucleus in both the dispersion
and spatial directions. Several techniques were used to remove these
effects.

Corrections for scattered light in the spectra were applied in the
following order: 1) removal of the reflection spectrum (in the CCD
spectra), 2) correction for the halo, and 3) removal of the remaining
PSF, including the diffraction-ring tracks. The reflection spectrum is
not only shifted in both directions, it is broadened in the spatial
direction, compressed in the dispersion direction, and altered in
intensity as a function of wavelength (it tends to be redder than the
nuclear spectrum).  The reflection in each original spectral image was
isolated by subtracting the scattered-light at the same spatial
distances on the other side of the nuclear spectrum.  Then the nuclear
spectrum was shifted along the slit and compressed in the dispersion
direction until the strong emission features matched those in the
reflection. It was then divided into the observed reflection to obtain
the large-scale intensity variations in both dispersion and spatial
directions. These variations were fitted in wavelength regions that do
not contain extended emission, in both directions with low-order
splines. The fits were then multiplied by the altered nuclear spectrum
to produce a model of the reflection which was subtracted from the
original spectral image. A circularly-symmetric halo was adopted from
previous work on the STIS detectors (Lindler 1999), and collapsed to
match the observed PSF in the spatial direction (obtained by adding
regions along the dispersion direction that do not contain extended
emission). The halo function was adjusted at various radial positions
until a reasonable match was obtained with the broad-scale profile of
the PSF (i.e., ignoring diffraction tracks, etc.). The halo was then
deconvolved from the original image using an iterative technique that
removes flux from the halo and places it in the core.

To remove the remaining scattered light, a scattering template was
constructed using archival observations of stars observed with the
same grating and slit width. First, the template spectrum was
normalized in the dispersion direction by dividing through by the
spectrum summed along the slit. Next, the template was smoothed in the
dispersion direction, using a median filter with a 50 pixel wide
window. The nuclear spectrum of NGC 4151 was then multiplied into the
template to simulate the scattered light spectrum. The scattering
subtracted spectra are clean of broad line emission as close as 4
pixels from the nucleus.  Because the nuclear \Ha line in the
G750L spectrum at P.A. 221 is saturated, the true line profile is
distorted and complicates construction of the scattering template.  A
substitute for the saturated profile was obtained from the G750M short
exposure in our slitless spectroscopy with good results.

An alternative approach was also applied which used the structure
along the slit in a continuum region of the NGC 4151 spectrum itself
to form the model template.  First, the entire image was normalized by
dividing each row (which lies along the dispersion direction) by the
summed nuclear spectrum from the central four rows. A spline
(typically of order 11) was then fitted along each row in regions that
do not contain emission lines.  Thus the fit is a model of the
scattering as a function of wavelength and position along the slit for
a point source spectrum of constant flux per unit wavelength.  This
procedure was effective in modeling the diffraction tracks as well as
the overall PSF.  The spline fits were then multipled by the nuclear
spectrum at each spatial position, and subtracted from the reflection-
and halo-corrected image to produce a final corrected image, which was
used for subsequent analysis.

\begin{figure}
\vspace*{2.0truein}
\hspace*{0.4truein}
\centerline{\LARGE FIGURE 2 --- 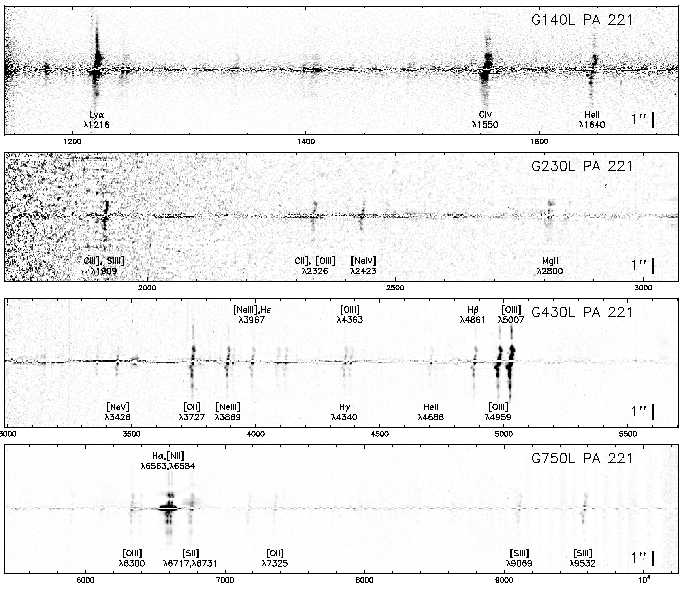}
\vspace*{2.0truein}
\figcaption{The fully reduced, scattering corrected images are
displayed for the observations taken at P.A. 221\degx. The dynamic
range displayed is the same for all frames and runs from zero for the
white background to $2.0 \times 10^{-16}$ ergs s$^{-1}$ cm$^{-2}$ \AA$^{-1}$ 
for black.  The receding NE side of the galaxy is at the top. 
\label{pa221}}
\end{figure}

\begin{figure}
\vspace*{2.0truein}
\hspace*{0.4truein}
\centerline{\LARGE FIGURE 3 --- 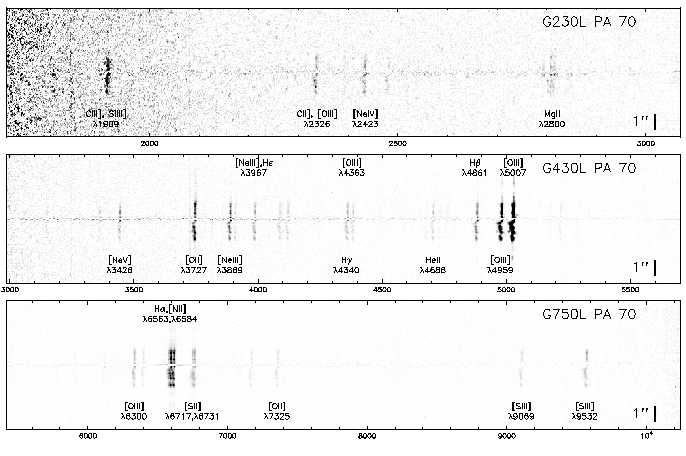}
\vspace*{2.0truein}
\figcaption{The reduced and scattering corrected images are displayed
for the observations taken at P.A. 70\degy.  The dynamic range
displayed is the same for all images and runs from zero for the white
background to $2.0 \times 10^{-16}$ ergs s$^{-1}$ cm$^{-2}$ \AA$^{-1}$ 
for black. The receding NE side of the galaxy is at the top. Note that
the G140L observations were not obtained for this orientation. 
\label{pa70}}
\end{figure}

The resulting spectra are shown in Figure \ref{pa221} and Figure
\ref{pa70}.  The corrections bring out the structure in the bright
lines, and allow us to see fainter lines that are not evident in the
original spectra. The correction process was not perfect, as evidenced
by the faint structure seen in the regions above and below the strong
nuclear lines, particularly \Ha $\lambda 6563$. However, these
problems are minor, and the contaminating effects of nuclear
absorption and emission were removed well enough for accurate
measurement of the extended emission, even very close to the nucleus.

Although our primary interest is the NLR, the data set also
contains high quality nuclear spectra of NGC 4151 at two epochs.
Monitoring campaigns have shown pronounced variability in both the
nuclear continuum and BLR emission (Robinson {\it et al.}, 1994,
Crenshaw {\it et al.}, 1996, Warwick {\it et al.}, 1996, Kaspi {\it et
al.}, 1996, Ulrich {\it et al.}, 1997, Weymann {\it et al.}, 1997,
Peterson {\it et al.}  1998).  Over a time interval of 33 days
(Jan. 8, 1998 and Feb. 10, 1998; see Table 1) the nuclear continuum
dropped by 17\% at 3050 \AA~ and 10\% at 6924 \AA~ decreasing
monotonically between these two wavelengths. This degree of variation
is consistent with that reported in short timescale variability
studies \cite {kaspi96}.  The variation of the BLR emission lines is
less pronounced than that of the continuum, with H$\gamma$ and \Hb
showing a decrease in flux, while the change in the \Ha + N[II] line
profile is more difficult to evaluate.  The absorption lines in our
far-UV spectrum are similar to those in the FOS spectra published by
Weymann {\it et al.}  (1997) but is at too low a resolution for
comparison to the high resolution GHRS spectrum.

\section{Analysis}
\label{ana}

\subsection{Measurement of Line Fluxes and Component Deblending}
\label{measure}

Emission line fluxes and their errors were measured along the slit in
each spectral range for a total of 45 emission lines.  Individual
spectra were extracted from the longslit spectra by summing along the
slit.  The size of the extraction bins was dictated by the need for
reasonably accurate fluxes for the He II $\lambda$1640 and
$\lambda$4686 lines, which were used for the reddening corrections in
Paper II. Experimentation revealed that bin lengths of 0\asecx.2 (4
CCD pixels, 8 MAMA pixels) within the inner $\pm$1$''$ and 0\asecx.4
outside this region would provide reasonable signal-to-noise ratios
for these lines, and still isolate the emission-line clouds that we
identified in our earlier papers. In some cases slightly different bin
sizes were used to isolate individual clouds or to increase the
signal-to-noise ratios.

To measure the line fluxes, first a linear fit to the continuum
adjacent to each line was subtracted.  Typically the continuum was
very close to zero following removal of the scattered light, but
continuum subtraction was helpful in regions of residual
structure. Next, the extreme ends of the red and blue wings of the
line were marked and the total flux and centroid were computed between
these two points. The uncertainties in the line fluxes were estimated
using the error arrays for each spectrum produced by CALSTIS and a
propagation of errors analysis (Bevington, 1969). For the blended
lines of \Ha and [N$\,$II]$\lambda\lambda$6548, 6584, and [S$\,$II] $\lambda
\lambda$ 6717, 6731, we used the [O$\,$III]$\lambda$5007 line as a
template to deblend the lines (see Crenshaw \& Peterson 1986). This
was superior to Gaussian fitting since the emission line profiles are
often complex. The results of the emission line flux measurements are
presented in Table 2 where the flux values are listed relative to \Hb
and the \Hb flux is given at the bottom in units of $10^{-15}$ ergs
cm$^{-2}$ s$^{-1}$ \AA$^{-1}$. The errors obtained for each flux are
given in parentheses.

Because of the failure of the far-UV spectrum at P.A. 70\degx no
G140L spectrum was obtained and so a reddening correction using the He
II lines was not possible.  Although dereddening using the Balmer
decrement is certainly valid, a reliable extrapolation from the red to
the blue and near-UV lines is uncertain. We prefer, therefore, to
continue the analysis without the corrected line ratios taking care
that any possible effects of reddening are accounted for by other
means.  Correction of the line ratios for reddening is an important
step for a detailed photoionization model and is therefore presented
in Paper II for the data at P.A. 221\degy.

To extract information on the multiple kinematic components the \Oiii
and \Oiiix lines were fitted independently with one to three
Gaussians.  Many slit extractions showed two components although in
only a few cases was there compelling evidence for a third.  Only
velocity components measured independently at both \Oiiix and
\Oiii are included.  To test that each component represented the true
kinematics of the gas, we compared the velocities obtained at both
\Oiii and [OIII] $\lambda $4959. Only those components with velocity
difference in the two lines less than or equal to twice the mean
difference for all points were retained. The procedure was then
repeated.  The first iteration removed velocity components that were
wildly discordant, and therefore unlikely to be real, while the second
gave us confidence that the remaining components are physically
significant. From the difference in velocity between components
extracted at \Oiiix and [OIII] $\lambda $5007, we estimate the standard
deviation of the velocities to be 30 \kms. The results for each slit
position are given in Table 3a and 3b where the Gaussian components
are listed in order of increasing velocity.  Negative slit
positions correspond to the SW region and positive slit positions
correspond to the NE.

\subsection{Kinematics}
\label{kinematics}

\begin{figure}
\vspace*{-1.4truein}
\hspace*{-0.75truein}
\plotonex{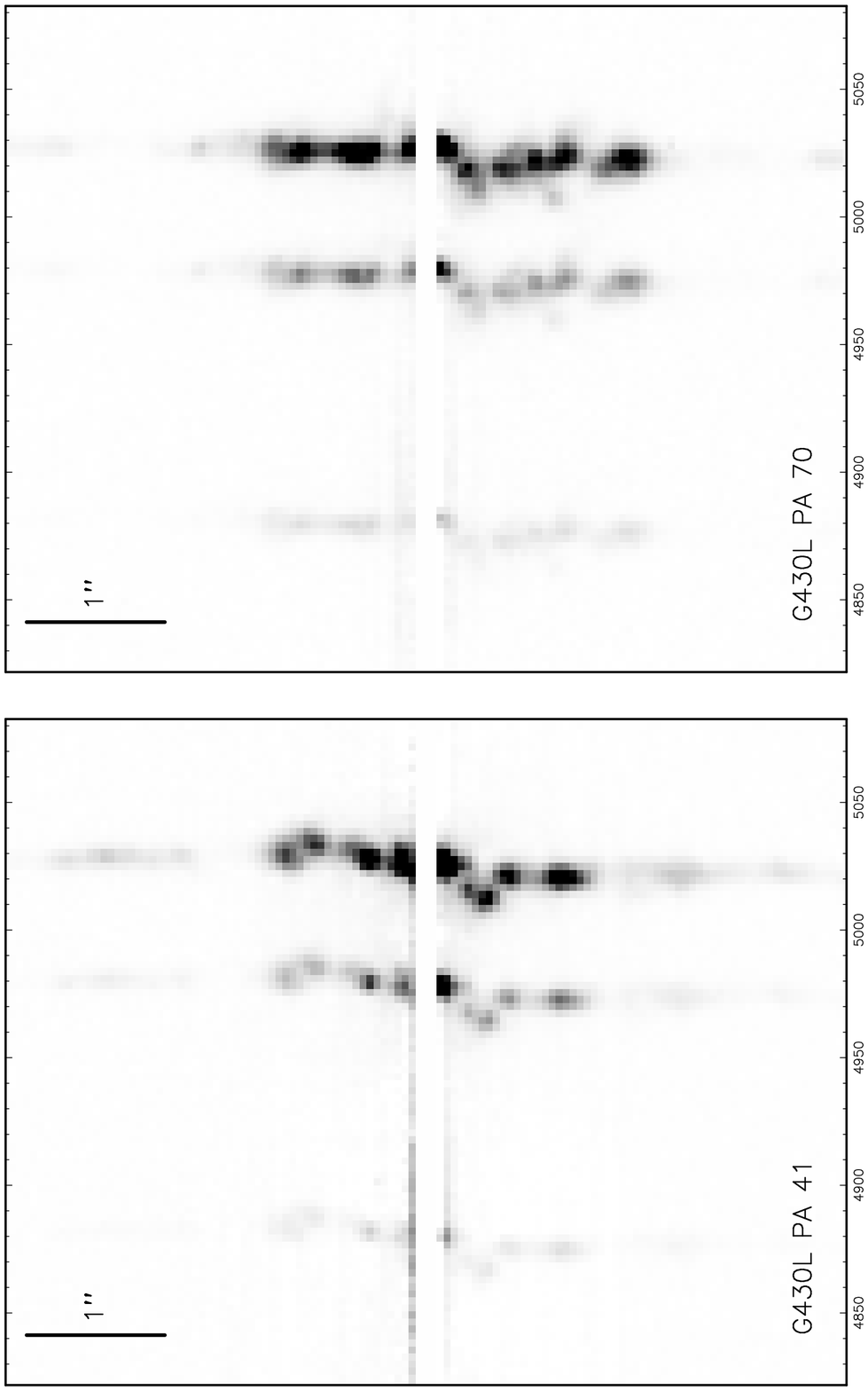}{6.5in}
\figcaption{Portions of the longslit spectra centered on the [OIII]
$\lambda $5007,\Oiiix and \Hb lines for each slit position. As in
Figures \ref{pa221} and \ref{pa70} NE is at the top.
\label{oiiihb}}
\end{figure}

Figure \ref{oiiihb} shows portions of the longslit spectra centered on
the \Oiii, \Oiiix and \Hb emission lines for both slit positions, with the NE
end of the slit at the top. The complex velocity structure that has been
noted in both ground-based and HST studies ({\it e.g.} Schulz 1990,
Kaiser {\it et al.} 1999, Hutchings {\it et al.}, 1999) is seen
including line splitting at several positions along the slit. Note
that as a result of our scattering correction and PSF subtraction we
are able to probe the emission line kinematics to within 0\asecx.2 of
the nucleus.

Included within our slits are four of the high velocity regions
(absolute value of projected velocity greater than 400 \kms) reported
in Hutchings {\it et al.} (1999) and 20 of the clouds identified in
Kaiser {\it et al.} (1999) (Tables 4a, 4b ). Our agreement with
Hutchings' velocities is reasonable, ranging from a difference of 6
\kms for region N detected at slit P.A. 221\degx, to 160 \kms for
region D.  While some of the difference is undoubtedly due to
measurement uncertainties, there may be real differences due to the
portion of the high velocity regions which fall within our slit.
There may also be some uncertainty due to confusion of spectral and
spatial information in the slitless data.  We see components of high
velocity gas not specifically identified by Hutchings {\it et al.}
(1999) on both sides of the bicone at both slit positions (Tables 3a,
3b). This gas corresponds to high velocity gas imaged by Hutchings
{\it et al.} (1999), but for which velocities were not previously
measured.  The high velocity components generally account for a small
fraction of the total flux in the [OIII] emission lines, again in
agreement with the findings of Hutchings {\it et al.} (1999). We find
more high velocity components in slit position P.A. 70\degy, which is
close to the radio ridge line, than we do in the P.A. 221\degx slit.
However, there is some high velocity gas not associated with the radio
emission.

Comparison of our velocities with those reported by Kaiser {\it et
al.} (1999) is more difficult, since in several cases they reported
single velocities for clouds where we find multiple velocity
components. Furthermore, there are instances of extended clouds for
which our slit does not sample the entire cloud.  If we compare only
velocities for clouds for which we find a single velocity component
and average our velocities for clouds occurring in more than one
extraction bin, we find the average difference in velocities is -18
\kms +/- 94 \kms, (in the sense V(this paper) - V(Kaiser {\it etal.})).  
This difference and range is comparable to what was found for
the high velocity clouds, and can be attributed to the same causes.

\begin{figure}
\vspace*{-1.0truein}
\hspace*{0.0truein}
\plotonex{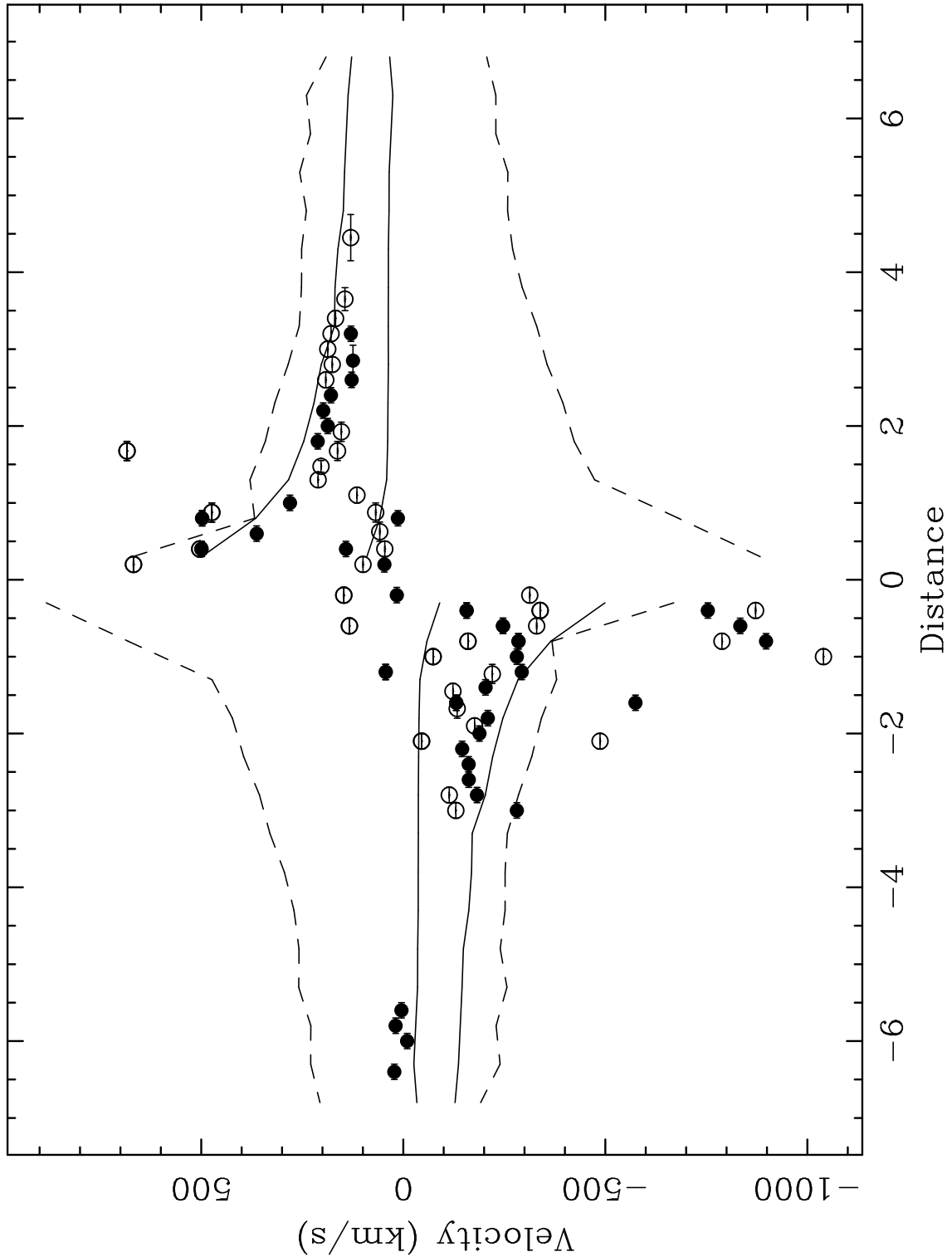}{6.5in}
\vspace{-0.5truein}
\figcaption{The velocities of the individual \Oiii components are
plotted as filled symbols for points along P.A. 221\degx and as open
symbols for points along P.A. 70\degx. The NE side of the slit is on
the positive side while the SW side corresponds to the negative
side. The uncertainties along the velocity axis are comparable to the
size of the points plotted and the horizontal bars indicate the bin
size of the extraction.  The solid line shows the expected velocities
in the case of a decelerating radial flow while the dashed line shows
the case of a decelerating expansion perpendicular to the radio axis.
\label{oiiivels}}
\end{figure}

Figure \ref{oiiivels} shows the velocities of the individual [OIII]
components from the Gaussian deblending.  Points along P.A. 221\degx
are marked as solid points and along P.A. 70\degx as open symbols. The
horizontal bars indicate the size of the extracted spectrum used for
the measurement along the slit. Vertical error bars are omitted since
the velocity uncertainties are comparable to the size of the points on
the diagram (see section \ref{ana}). A systemic velocity of 997 \kms
has been subtracted from the data. The solid and dashed lines show
results expected for our simple models described below.  The results
follow the velocity distribution determined from the slitless
spectroscopy of Kaiser $et~al.$ (1999) and the plot is similar to
their Figure 8, though without the extreme high velocities.  The
velocities at large distances from the nucleus are consistent with the
rotation of the galactic disk, while closer in the velocities are
strongly blue shifted SW of the nucleus and strongly redshifted to the
NE.

\begin{figure}
\vspace*{-1.50truein}
\hspace*{0.25truein}
\plotonex{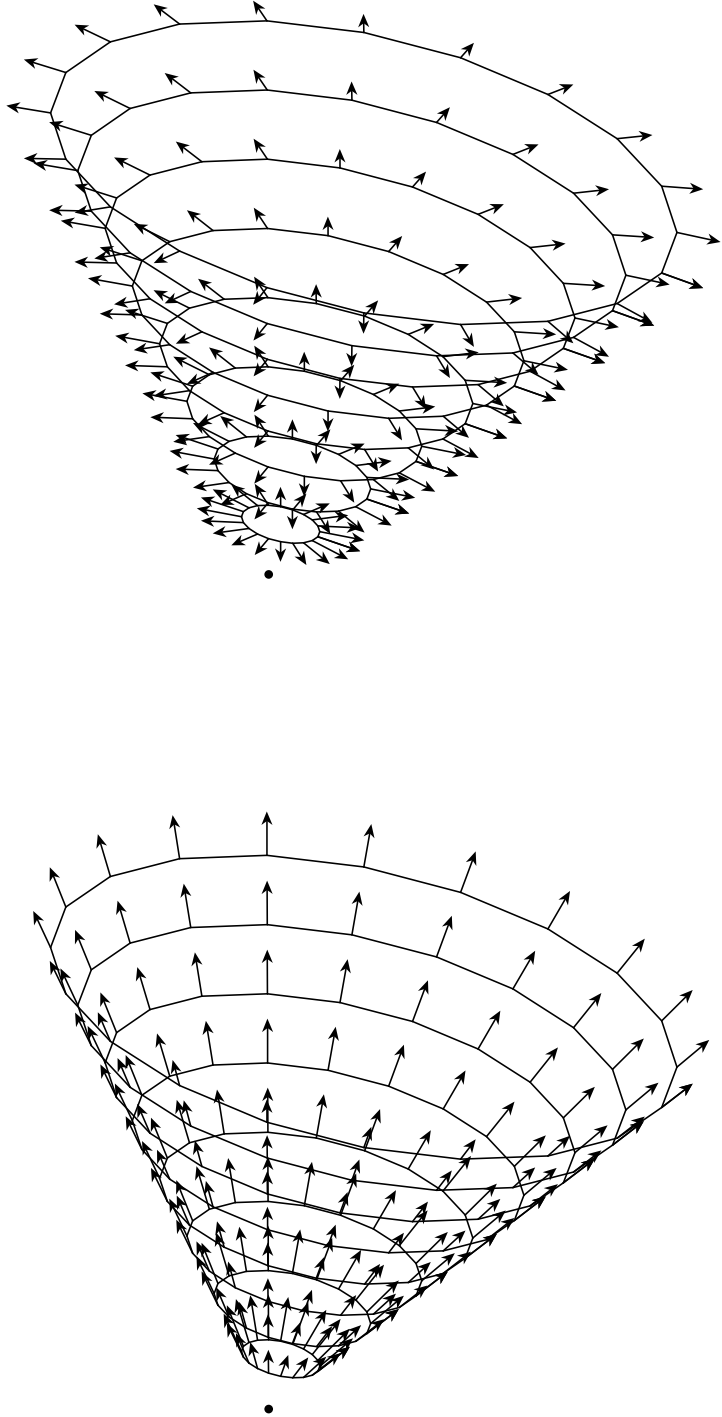}{6.5in}
\vspace{-0.50truein}
\figcaption{Examples of the model velocity fields are shown. We adopt
the geometry of Pedlar {et al.} (1993), however our opening angle is
smaller (70\degy) to better match the observed distribution of line
emitting material. The the upper cone shows the case of a flow
expanding in a direction perpendicular to the radio axis while the lower
cone shows the case of a radial flow. Velocity vectors are shown for
the outside surface of the cone.
\label{cones}}
\end{figure}

To better understand the kinematics we consider two possibilities for
the general form of the velocity field: radial outflow from the
nucleus and expansion directed away from the radio axis.  We adopt the
basic conical geometry of the NLR of NGC 4151 as modeled by Pedlar
{\it et al.} (1993), with the radio jet pointing 40\degx from the line
of sight and projected onto the plane of the sky at a P.A. of 77\degy.
After consideration of the well-known geometry of the host galaxy
(Simkin, 1975, Bosma, Ekers, \& Lequeux, 1977) we require that the
cone opening angle be wide enough to include our line of sight to the
nucleus and also to intersect the disk of the host galaxy, since the
Extended Narrow Line Region (ENLR) kinematics follow the rotation of
the disk. Pedlar {\it et al.} (1993) estimate the opening angle to be
130\degy.  However, Boksenberg {\it et al.}  (1995) argue that the NLR
is density bounded and the ionized gas only partially fills the cone.
Therefore we choose a narrow vertex angle of 70\degx which is a better
match to the observed NLR structure.  The models are drawn
schematically in Figure \ref{cones}. These models are used to estimate
the radial velocity as a function of projected distance from the
nucleus for each slit position angle.  Our purpose is not to
produce a detailed match to the observed velocities of each individual
cloud, but to test two ideas about general form of the NLR kinematics.
Therefore, we assume that the interior of the cone is uniformly filled and
note that the observed velocity distribution is not expected to be as
smooth or complete as the model, reflecting the way in which the
emission line clouds fill the cone.

\begin{figure}
\vspace*{2.0truein}
\hspace*{0.25truein}
\centerline{\LARGE FIGURE 7 --- 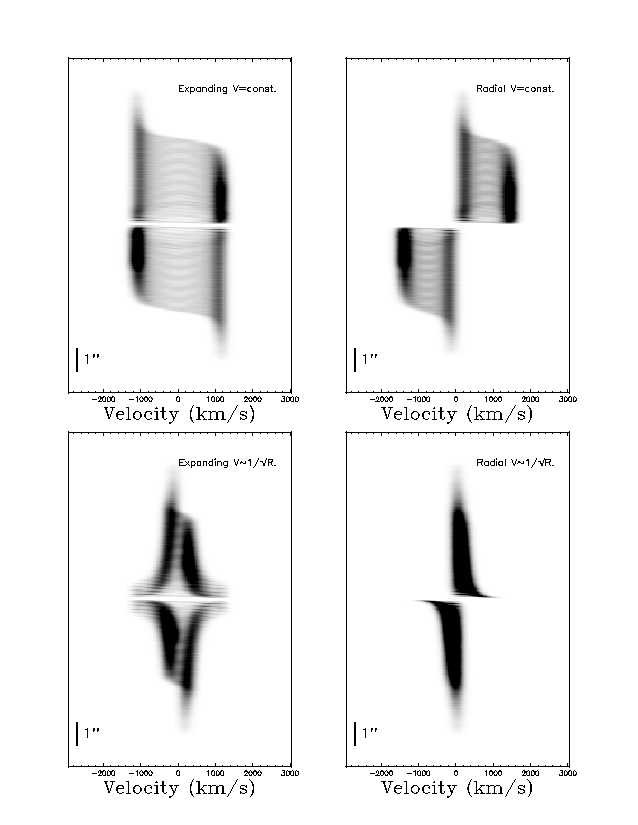}
\vspace{2.0truein}
\figcaption{Our models can be used to produce artificial longslit
emission line spectra. On the left we show a model emission line
for the case of expansion away from the radio axis and on the right for
a radial flow away from the nucleus. Models with constant velocity 
are shown at the top and for decelerating flows on the bottom.
\label{rvmod}}
\end{figure}

For both the radial outflow model and the jet expansion model we
consider two cases, one in which the flow has a constant velocity and
one in which the flow decelerates as it moves outward. We model this
decelerating flow as a $R^{-1/2}$ dependence where $R$ in the radial
flow model is distance from the nucleus and $R$ in the jet expansion
model is distance from the radio axis. This particular form of
deceleration is chosen since it seems to represent the data best and
is meant only to illustrate the effect. The results are plotted in
Figure \ref{rvmod} for all four models in the form of a model longslit
spectrum of a single emission line comparable to Figure
\ref{oiiihb}. The slit was chosen to lie along P.A. 70\degx and to
have a slit width of 0\asecx.1 as in our STIS observations. These
simulated spectra were then deblended using two Gaussians at each slit
position in the same manner as the real data.  The velocities for the
decelerating models are shown in Figure \ref{oiiivels} as the dashed
lines (one for each Gaussian component) for the case of jet expansion
and the solid lines for the case of radial flow.

For the case of expansion away from the radio axis we expect both
large positive and large negative velocities relative to the systemic
velocity at any given position along the slit.  In the case of radial
outflow, however, large positive velocities and velocities much closer
to the systemic velocity will be observed on one side of the slit
while on the other side, large negative velocities and velocities near
the systemic velocity are expected.  Since for NGC 4151 the far side
of the SW cone lies close to the plane of the sky, the bulk of the
flow is transverse to the line of sight, yielding radial velocities
close to the systemic value, while the near side is much closer to the
line of sight yielding large approaching radial velocities. Similar
considerations hold for the NE cone except that the near side of the
cone lies in the plane of the sky and the far side yields the large
receding velocities.

We conclude that the radial outflow case gives a better match to the
observed velocity distribution than the case of expansion away from
the jet. In the case of a radially decelerating flow the overall
envelope of the highest velocities decreases as one moves away from
the nucleus much as seen in Figure \ref{oiiivels}. Although the match
is not perfect it seems to follow the trend of less extreme velocities
as one moves along the slit.  From these simple models we cannot
exclude the possibility of some motion perpendicular to the radio
jet. However, it does seem likely that the flow is dominated by a
radial outflow from the nucleus which slows with distance and that any
contribution from expansion away from the jet is less significant.

\subsection{Line Ratios and Projected Distance from the Nucleus}
\label{ratdist}

An understanding of the physical conditions in the NLR can be
obtained by considering how various line strengths change as a
function of distance from the nucleus and with respect to each other.
In Paper II (Kraemer {\it et al.} 1999) a detailed photoionization model
is developed using the emission line fluxes presented here. In the current 
paper we present a simpler analysis.

\begin{figure}
\vspace*{-1.5truein}
\hspace*{0.25truein}
\plotonex{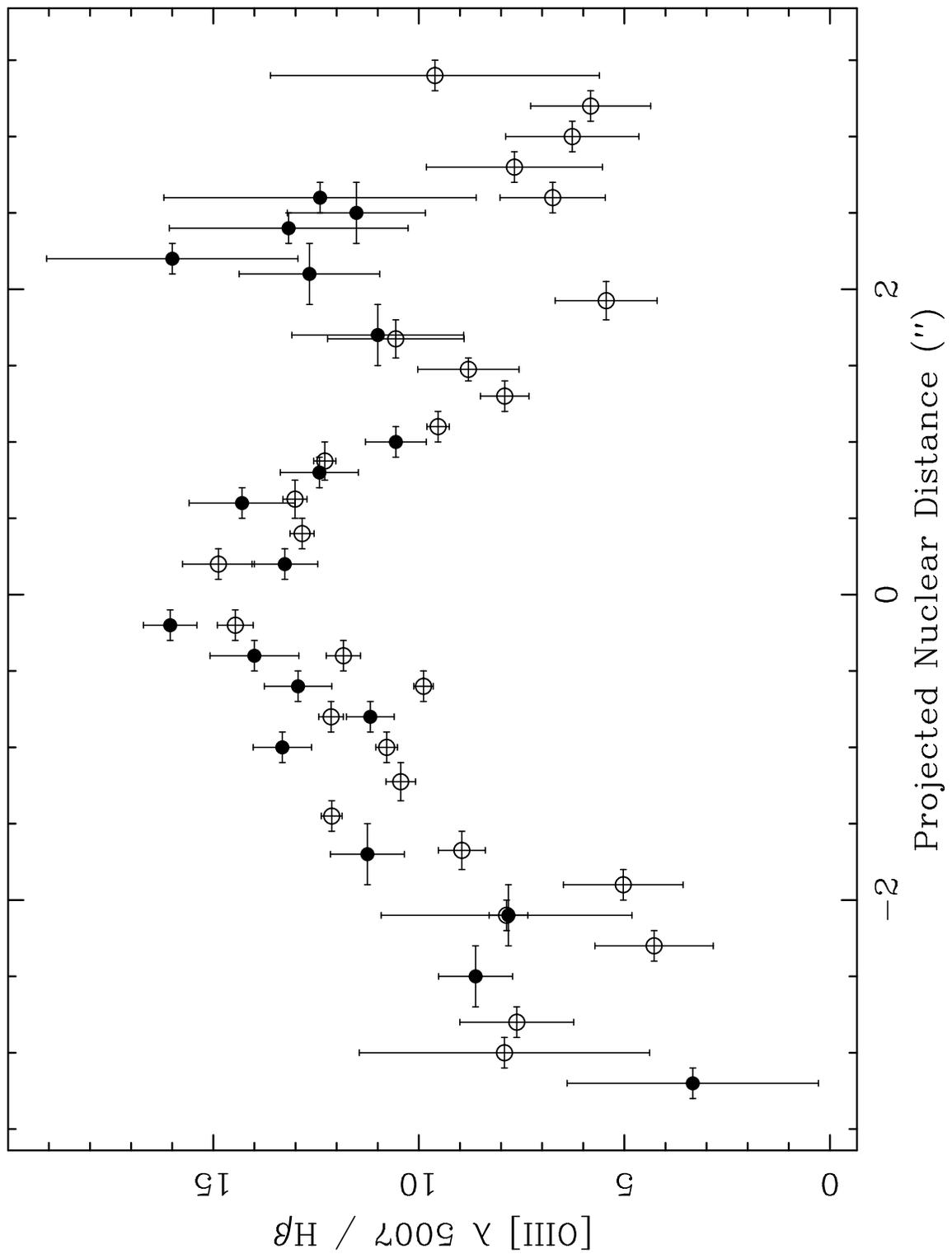}{6.5in}
\vspace{0.0truein}
\figcaption{The ratio of \Oiii to \Hb is plotted as a function
of distance from the nucleus along the slit. Notice the trend for the ratio
to decline with distance in the inner 2\asec.
\label{o3hbrat}}
\end{figure}

The ratio of \Oiii to \Hb is well known to be sensitive to the
ionization parameter $U = Q / 4\pi r^2 n_e c$, where $Q$ is the rate
at which ionizing photons are emitted, $r$ is the distance to the
nucleus, $n_e$ is the electron density, and $c$ is the speed of light.
Figure \ref{o3hbrat} shows the \Oiii to \Hb ratio as a function of
distance along the slit for both position angles. We use ratios that
have not been corrected for extinction since the lines are close in
wavelength and are therefore rather insensitive to reddening. We see
from the diagram that the line ratio decreases with distance in the
inner 2\asec and recovers somewhat at larger radii on the NE side
(positive X-axis). This trend was shown in Kaiser {\it et al.} (1999)
from the slitless spectroscopy who suggest that this apparent change
in the ionization parameter with distance reflects a decrease in
density. 

Apart from the increase in \Oiii / \Hb on the extreme NE side of
P.A. 221\degx, there is no significant difference in the ratio between
the two position angles, suggesting that while the ionization state of
the gas may change moving away from the nucleus, it generally does not
change laterally, {\it i.e.} with distance from the radio axis.

\begin{figure}
\vspace*{-1.5truein}
\hspace*{0.25truein}
\plotonex{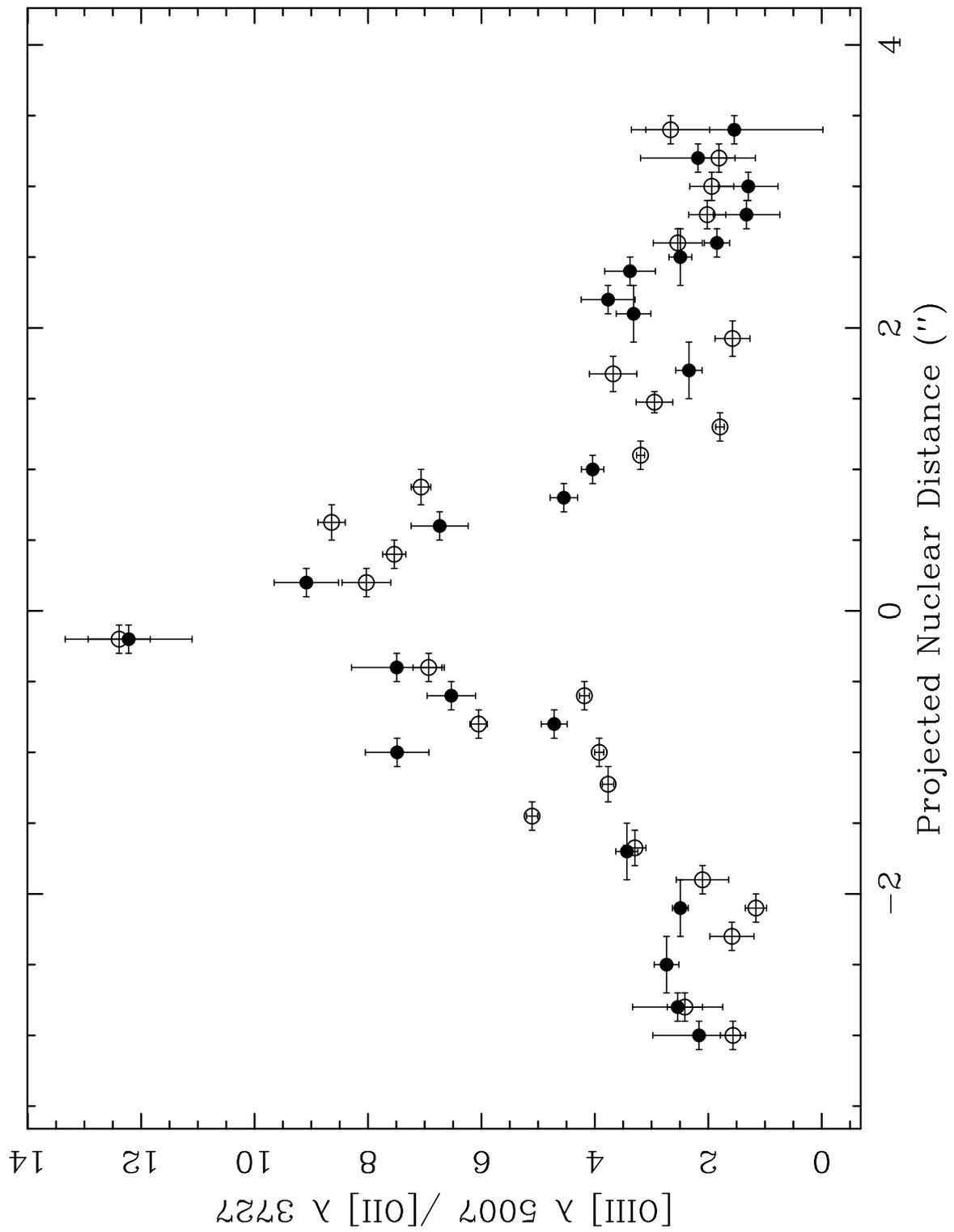}{6.5in}
\vspace{0.0truein}
\figcaption{The ratio of \Oiii to \Oii is plotted as a function
of distance from the nucleus along the slit. 
\label{o3o2rat}}
\end{figure}

A similar trend is seen in the ratio of \Oiii to \Oii, which is also
sensitive to the ionization parameter. Figure \ref{o3o2rat} plots the
ratio versus distance for both slit positions.  Again, the line fluxes
have not been corrected for extinction but the dust is most likely
patchy (see Paper II) and so is not likely to influence the overall
trend and merely adds scatter. Support for this comes from the fact
that the trend is largely symmetric about the nucleus indicating that
no large scale dust lanes pass through our aperture.  Furthermore, for
the slit extractions where both He II lines used for the extinction
corrections are present (along P.A. 221\degx), the largest change in
the [OIII]/[OII] ratio from dereddening was a decrease of $\sim 30$\%
(see Paper II).  Therefore, to the extent that the distribution of
dust is comparable along each slit position, the conclusion of a
decreasing [OIII]/[OII] ratio with distance is robust.

The safest conclusion to draw from these diagrams is that the density
falls off with distance as suggested by Kaiser {\it et al.} (1999) and
confirmed in Paper II. In fact, in the inner clouds the high
[OIII]/[OII] most likely results from collisional de-excitation of the O$^+$
ions. Although, these trends could naively be considered an indication
of decreasing ionization parameter with distance from the nucleus,
the more detailed investigation in Paper II suggests a more constant
ionization parameter and a density which declines less rapidly than
$r^{-2}$.

\begin{figure}
\vspace*{-1.5truein}
\hspace*{0.25truein}
\plotonex{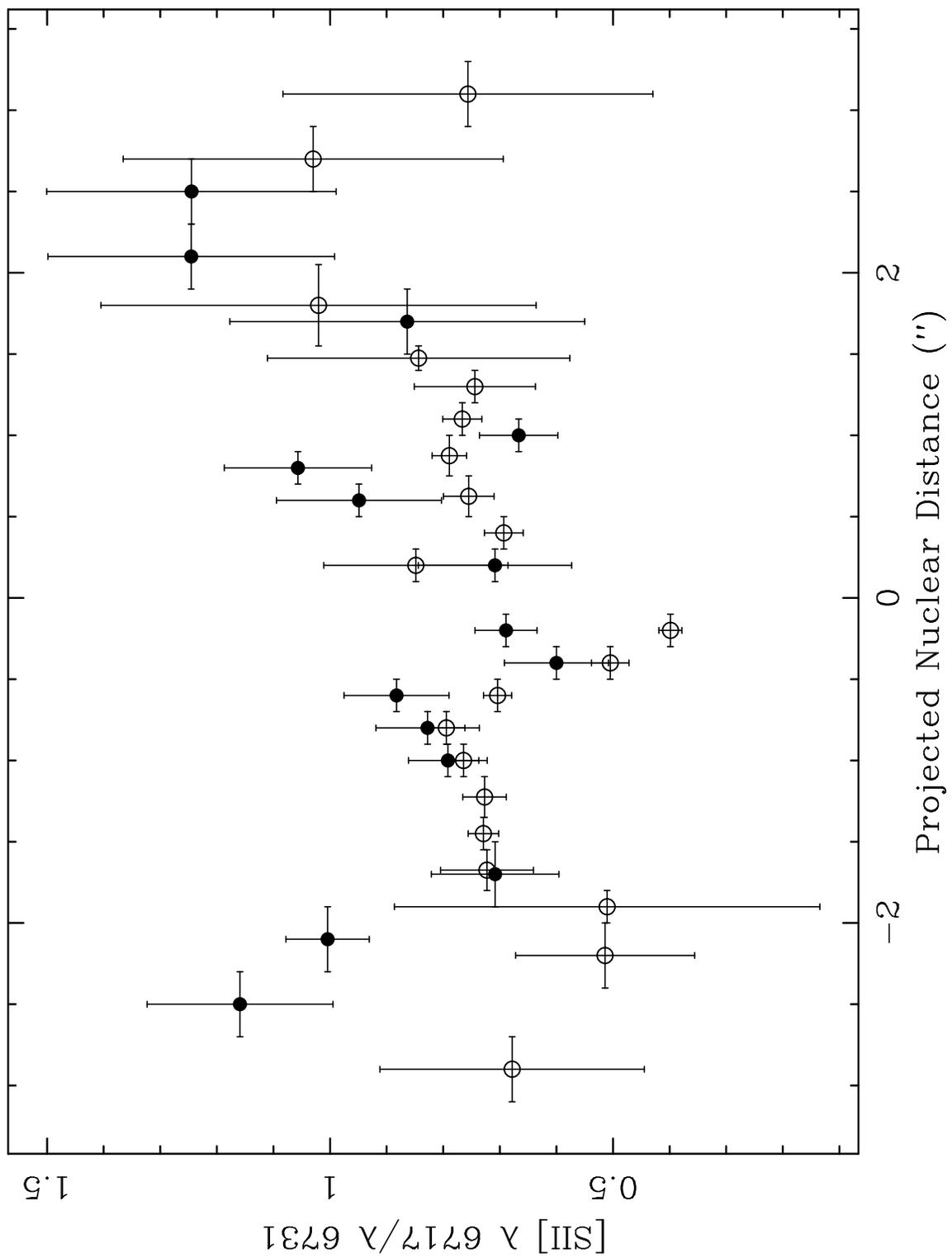}{6.5in}
\vspace{0.0truein}
\figcaption{The density sensitive ratio \Siia/\Siib is plotted as a function
of distance from the nucleus along the slit.
\label{s2vdist}}
\end{figure}

In Figure \ref{s2vdist} the density sensitive ratio of \Siia to \Siib
is plotted as a function of distance, again for both slit positions.
Judging by the size of the error bars, much of the scatter in the
diagram is real suggesting that the gas is rather clumpy, with regions
of higher and lower density at various points along the slit.  There
is also an interesting drop in the ratio very close to the nucleus
particularly in the data from the P.A. 70\degx slit position,
suggesting an increase in density there.  Generally, the [SII] ratios
appear to be larger farther out particularly along P.A. 221\degx
(solid dots), indicating a decrease in density with radius, at least
in the partially ionized zone.  Using the five level-atom program
developed by Shaw \& Dufour (1995) and assuming a temperature of 15000
\degx K (see below) we find that the density of the inner NLR is
roughly $2000 \rm ~cm^{-3}$ while in the outer NLR and ENLR the
density has dropped to $\sim 300 \rm ~cm^{-3}$. This agrees with the
results of Robinson {\it et al.}  (1994) who found density decreasing
with distance from the nucleus in NGC 4151, with an overall NLR
density of $1600~ \rm cm^{-3}$, and a density in the ENLR of $250~ \rm
cm^{-3}$. This is also in agreement with the interpretation of the
decline in \Oiii/\Hb as the result of a decrease in density.

\begin{figure}
\vspace*{-1.5truein}
\hspace*{0.25truein}
\plotonex{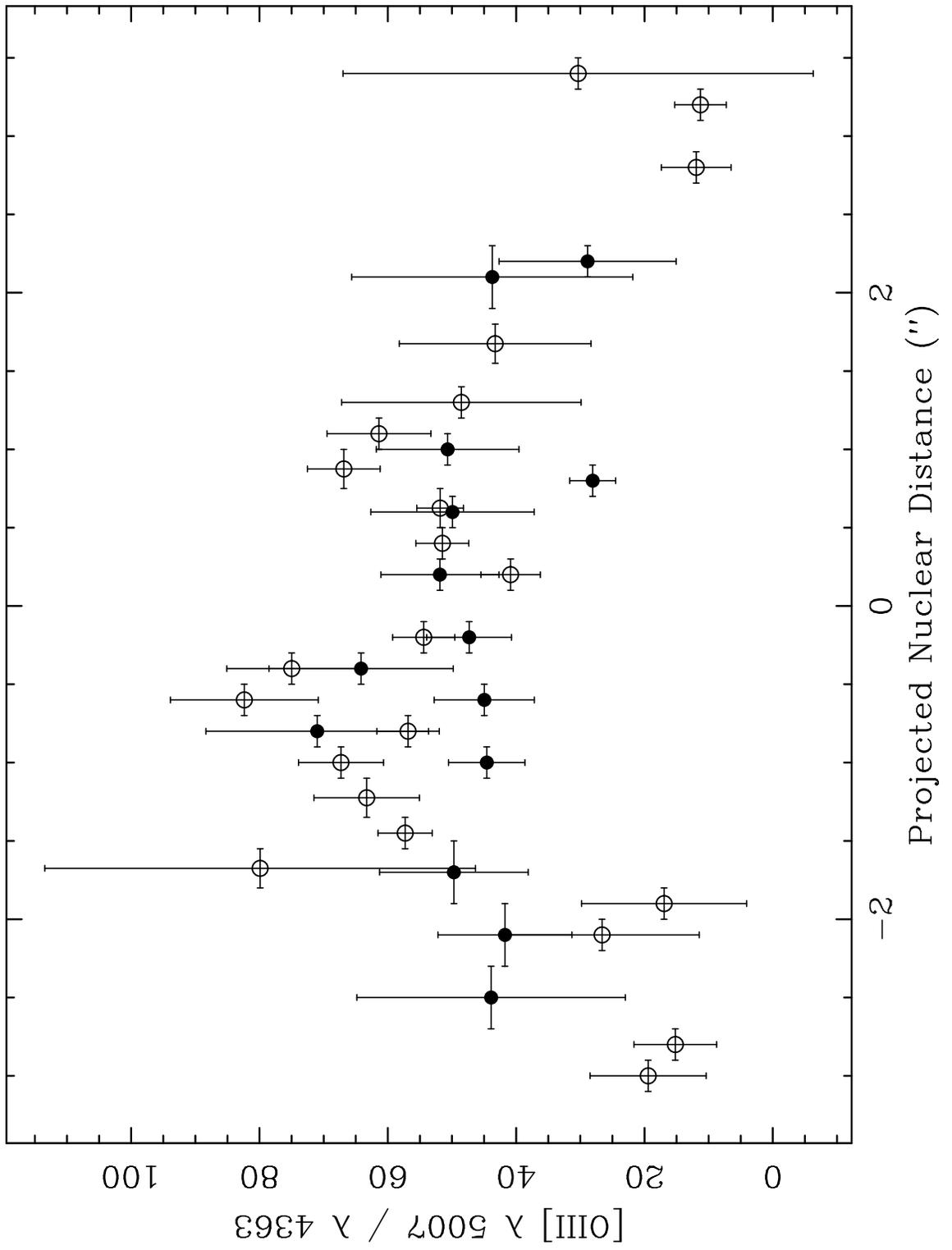}{6.5in}
\vspace{0.0truein}
\figcaption{The temperature sensitive \Oiii/ [OIII] $\lambda 4363$ ratio
is plotted as a function of distance from the nucleus along the slit.
\label{o3o3vdist}}
\end{figure}

The [OIII]$\lambda 5007$/[OIII]$\lambda 4363$ ratio is well-known to
be sensitive to the temperature of the gas. Figure \ref{o3o3vdist}
shows the [OIII] ratio as a function of distance along the slit.  The
use of this ratio to calculate the temperatures is only valid for
densities up to $n_e \simeq 10^5$ cm$^{-3}$ at which point collisional
de-excitation begins to have an influence on the line strengths
(Osterbrock, 1974).  Furthermore the [SII] densities cannot be used
since they reflect densities in the partially ionized zone. Thus we
use results from Paper II for the gas densities which indicate that in
the O$^{++}$ zone the densities are below $\rm 10^5 cm^{-3}$. The
results from the five-level atom program give temperatures in the
range of 12000\degx K --- 17000 \degx K. Based on Figure
\ref{o3o3vdist} there appears to be a slight trend for a decreasing
ratio (increasing temperature) with distance from the nucleus. This is
difficult to confirm, however, since reddening may play a role,
tending to increase the observed ratio. Paper II gives a more detailed
analysis of the physical conditions along the slit.

\subsection{Line Ratio Diagrams and Photoionization}
\label{ratdiags}

Diagrams plotting one line ratio against another can be used to
investigate the origin of the photoionizing continuum. By choosing
line ratios which consist of lines which are close in wavelength we
can significantly reduce the effects of reddening (see {\it e.g.}
Veilleux and Osterbrock, 1989). In Figure \ref{ratioplots} a, b, and
c, we present the optical emission line ratios [S II] $\rm \lambda
\lambda 6717,6731/H\alpha $, [N II] $\rm \lambda 6584/H\alpha $, [O I]
$\rm \lambda 6300/H\alpha $, respectively, plotted against \Oiii/\Hb.
In each diagram the solid line separates star-forming regions from AGN
and is taken from Veilleux and Osterbrock (1989). The dashed line is
the power-law photoionization model for solar abundance taken from
Ferland and Netzer (1983). The ionization parameter varies from
$10^{-4}$ to $10^{-1.5}$ from lower right to upper left.

We find that the NGC 4151 NLR clouds occupy compact regions on these
diagrams indicating that the source of the ionizing continuum is the
same for all of the points sampled along the slit. Thus none of the
clouds observed shows evidence for star-formation or
LINER-like excitation. While this result is not unexpected it is worth
commenting that the NLR gas all seems to have the same source of
excitation.

Other line ratio diagrams including UV lines are also interesting
since they allow us to investigate the possibility of alternate
ionization mechanisms for the NLR clouds (Allen {\it et al.} 1998).
In figure \ref{shockgrids}a, b and c we plot the ratios of CIV
$\lambda 1549$ to He II $\lambda 1640$, CIV $\lambda 1549$ to CIII]
$\lambda 1909$, [Ne V] $\lambda 3426$ to [Ne III] $\lambda 3869$,
respectively against \Oiii to \Hb (only the P.A. 221\degx data is
shown for Figures \ref{shockgrids}a and b since the far-UV observation
at P.A. 70 was unsuccessful). The lines show model grids calculated
using the MAPPINGS II code (Sutherland and Dopita, 1993) by Allen {\it
et al.}  (1999) for shock ionization (bottom), shock plus ionized
precursor gas (middle) and for power-law photoionization (top). For
the shock plus precursor models, the shock velocity increases from 200
\kms to 500 \kms moving from low to high \Oiii/\Hb ratios. Notice that
for the highest velocity shocks the models coincide with power-law
photoionization models.

Again the NGC 4151 NLR occupies very limited regions in these diagrams
corresponding to photoionization by a power law at high ionization
parameter or by shock plus precursor models with very high velocity
($V_{\rm shock} \simeq 500$\kms). These results strongly suggest that
low velocity shocks play an insignificant role in accounting for the
ionization state of the NLR in NGC 4151 but we cannot rule out the
possibility of ionization by radiation from fast shocks.

\section{Discussion}
\label{disc}

The results of the kinematic and emission line ratio analysis can be
combined to create a coherent picture of the NLR in NGC 4151. We have
seen that the kinematics bear the signature of radial outflow from the
nucleus and are distinctly different from an expansion away from the
radio jet axis. This is an interesting result since many recent
studies have reported kinematic evidence that the radio jet can have a
significant influence on the motion of NLR gas ({\it e.g.} Bicknell
{\it et al.} 1998, NGC 4151 Winge {\it et al} 1998, Mrk 3 Capetti {\it
et al.} 1998).  In these studies the NLR gas is found immediately
surrounding and expanding away from knots of radio emission as in Mrk
3 or forms a bow shock structure around the working surface of the
head of the jet as in Mrk 573 (Capetti {\it et al.}  1996, Falcke,
Wilson, \& Simpson 1998). This seems not to be the case for NGC 4151.

In conflict with this statement, the study of NGC 4151 by Winge {\it
et al.}  (1998) reports that high velocity clouds are seen around the
edges of the radio knots. This is not confirmed by Kaiser {\it et al.}
(1999) who conclude that there is no direct association between
non-virial gas kinematics, as determined by high velocity and high
velocity dispersion, and proximity to the radio knots. Our results
concur with those of Kaiser {\it et al.} (1999). While we do find high
velocity clouds in our aperture there is no distinct preference for
them to found along P.A. 70\degy, which is more closely aligned with
the radio axis (P.A. 77\degy).

Further support for radial outflow comes from the emission line ratios
as a function of position. For example there is no significant
difference in the [OIII]/[OII] or \Oiii/\Hb ratios between the two
slit positions even though the spectra at P.A. 70\degx are much more
closely aligned with the radio axis than the clouds at
P.A. 221\degy. This is in contrast to the case of NGC 1068 where the
[OIII]/[OII] ratio increases dramatically in regions that coincide
with the radio jet (Axon {\it et al.} 1998).  WFPC2 images of NGC 1068
presented by Capetti, Axon, \& Macchetto (1997) may also indicate
higher density and ionization state along the radio jet in this
object. Furthermore, these authors suggest that an additional source
of local ionizing continuum is required to explain the
observations. While these results certainly raise an interesting
possibility for NGC 1068, our results for NGC 4151 show no such
association between the radio morphology and the emission line ratios.
Thus the radio jet in NGC 4151 seems to have little influence on the
ionization state of the gas. Similar results are seen for the [SII]
ratio and the [OIII] $\lambda 5007/\lambda 4363$ ratio suggesting no
strong changes in the physical condition of the gas with proximity to
the radio emission.

Because the line ratio diagrams show no evidence for shock or shock
plus precursor ionization models at least for low velocity shocks,
they support the arguments for radial outflow.  If the gas were
expanding away from the radio axis one would expect to see large
amounts of shocked material particularly at the interface of the flow
with the ambient interstellar medium of the host galaxy disk.  In the
case of radial outflow, we would expect to see little shocked gas
since the motion is not directed into the disk and the relative
velocities of gas within the flow should be small.

Perhaps an important consideration is that the radio morphology of NGC
4151 is rather different from that of Mrk 3 for example. Pedlar {\it
et al.}  have compared the radio structure of NGC 4151 to that of an
FR I type radio galaxy, with much of the radio emission coming from a
diffuse component, although on much smaller scales. The radio emission
in Mrk 3, by contrast, is more jet-like being unresolved with MERLIN
perpendicular to the radio source axis (Kukula {\it et al.}
1993). Thus we might consider that the radio emission in NGC 4151 is
not a well collimated jet, but rather a broad spray of plasma. Gas
clouds in the vicinity of the radio flow would thus be more naturally
accelerated in directions roughly aligned with the radio axis than
perpendicular to it. 

One possible scenario is that the core of the radio jet in NGC 4151
has cleared a channel in the line emitting gas and has blown out of
the disk of the galaxy as suggested by Schulz (1988).  Thus there may
have been a bow shock associated with the radio lobes in the past but
the jet has passed on to lower density region in the outer bulge and
galaxy halo. The line emitting gas is now free to flow out along the
radio axis but only weakly interacts with the jet itself and the host
galaxy ISM.

NGC 4151 is also known to have a system of nuclear absorption lines,
particularly CIV $\lambda$1549, which are blueshifted with respect to
the systemic velocity by values ranging from 0 to 1600 \kms ({\it
e.g.} Weymann {\it et al.} 1997). It is tempting to link the outflow
seen in our study with that for the absorption line system.  However,
these flows are observed on vastly different scales and thus a true
connection has not been established. Models invoking winds from the
nucleus to explain the NLR kinematics and other properties of Seyfert
galaxies have been proposed ({\it e.g.} Krolik \& Vrtilek, 1984,
Schiano, 1986, Smith, 1993). One suggestion is that X-ray heating of
the molecular torus is the source of the wind (Krolik \& Begelman,
1986). The base of the wind forms the electron scattering region which
serves as the ``mirror'' allowing a view of the BLR in polarized light
in some Seyfert 2 galaxies. At larger radii one might expect that the
steep potential of the galaxy bulge tends to decelerate the wind.  We
conclude that the kinematics in NGC 4151 seem to be consistent with
wind models for the NLR.

\section {Summary}
\label{conc}

The results presented in this paper provide an interesting contrast to
the recent work on the NLR of Seyfert galaxies. Our analysis of the
longslit spectra of NGC 4151 has revealed a rather different picture
of the NLR in the sense that the prominent radio jet has very little
influence on the kinematics and physical conditions. We find that the
kinematics are best characterized by a decelerating radial outflow
from the nucleus in the form of a wind. The lack of evidence for
strong shocks near the radio axis and the uniformity of the line
ratios across the NLR supports this picture. Thus it appears that
while interaction between the radio jet and the NLR gas may be a common
occurrence it is by no means ubiquitous and does not apply in the case 
of NGC 4151.

We would like to thank Diane Eggers for her assistance in the data
analysis.  We would also like to thank Mark Allen for providing the
model grids for the UV line ratio diagrams.  This research has been
supported in part by NASA under contract NAS5-31231.

\clearpage
\onecolumn

\begin{table}

\begin{tabular}{llllllcrc}
\multicolumn{9}{c}{\bf TABLE 1} \\
\multicolumn{9}{c}{\bf STIS Longslit Observations of NGC 4151} \\
\multicolumn{9}{c}{~} \\
\hline\hline\\
& Dataset & Date & Grating & Dispersion & Spectral & PA~ & Ex. T. & \\
&  &  & & \AA/pixel & Range &  (\degy)  &   (sec)    & \\
\hline\\
& o42302070 & Feb. 10, 1998 & G430L & 2.73 & 2900$-$5700 & 221 & 720 & \\
& o42302080 & Feb. 10, 1998 & G750L & 4.92 & 5240$-$10270 & 221 & 720 & \\
& o42302090 & Feb. 10, 1998 & G140L & 0.6 & 1150$-$1730 & 221 & 2250 & \\
& o423020a0 & Feb. 10, 1998 & G140L & 0.6 & 1150$-$1730 & 221 & 2250 & \\
& o42303050 & Jan. 8, 1998  & G430L & 2.73 & 2900$-$5700 & 221 & 720 & \\
& o42303060 & Jan. 8, 1998  & G750L & 4.92 & 5240$-$10270 & 221 & 720 & \\
& o42303070 & Jan. 8, 1998  & G230LB & 1.35 & 1680$-$3060 & 221 & 2160 & \\
& o42304090 & June 1, 1998 & G430L & 2.73 & 2900$-$5700 & 70 & 720 & \\
& o423040a0 & June 1, 1998 & G750L & 4.92 & 5240$-$10270 & 70 & 720 & \\
& o423040b0$^{\ast}$ & Jun. 1, 1998 & G140L & 0.6 & 1150$-$1730 
   & 70 & 2250 & \\
& o423040c0$^{\ast}$ & Jun. 1, 1998 & G140L  & 0.6 & 1150$-$1730 
   & 70 & 2250 & \\
& o42305050 & May 7, 1998 & G430L  & 2.73 & 2900$-$5700 & 70 & 720 & \\
& o42305060 & May 7, 1998 & G750L  & 4.92 & 5240$-$10270 & 70 & 720 & \\
& o42305070 & May 7, 1998 & G230LB & 1.35 & 1680$-$3060 & 70 & 2160 & \\
& & & \\
\hline\\
&\multicolumn{8}{l}{$^{\ast}$ Observations failed.}\\
& & & \\

\end{tabular}
\end{table}

\clearpage

\begin{table}
\vspace*{-2.0truein}
\hspace*{-0.75truein}
\psfig{file=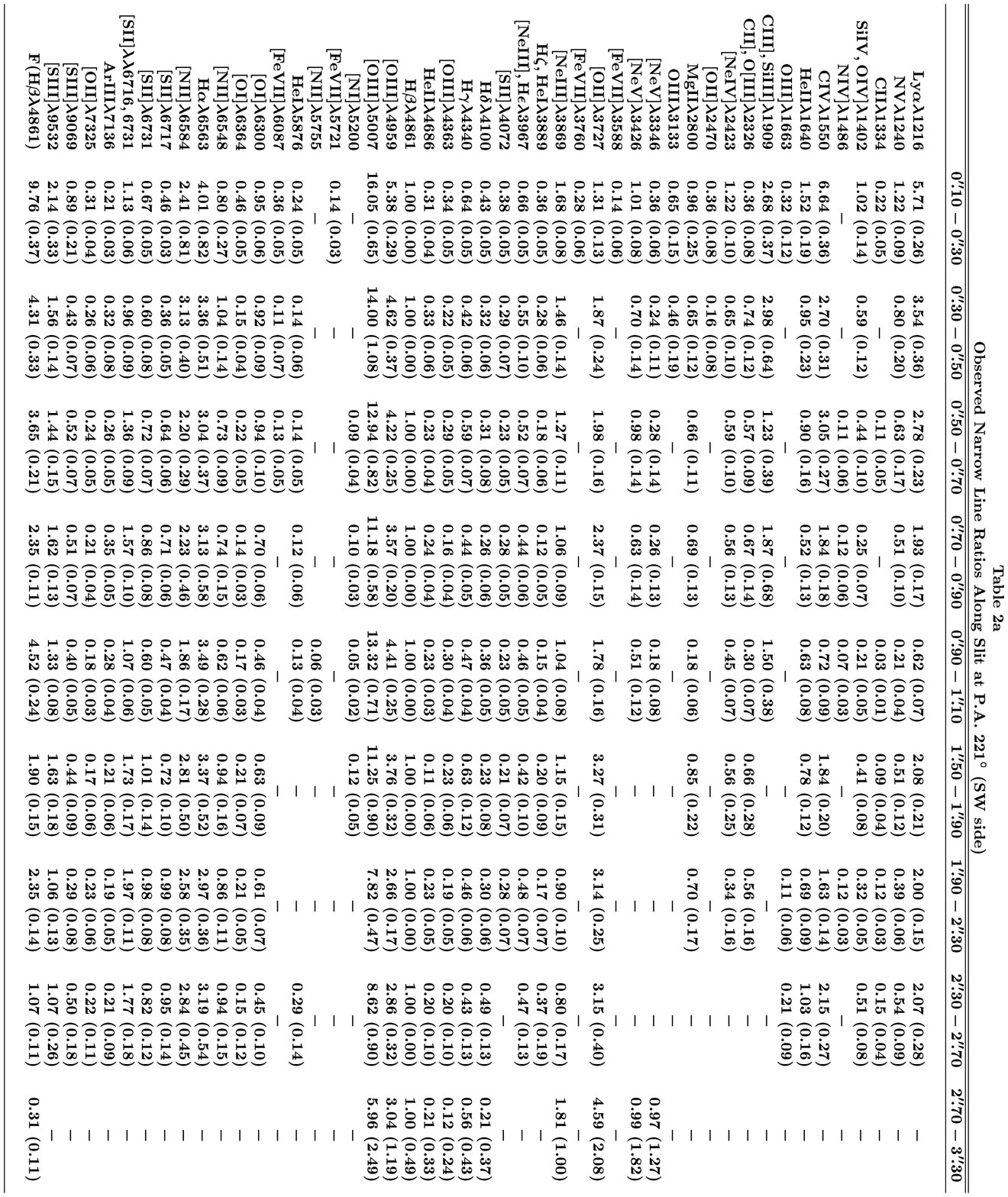}
\end{table}
\clearpage
\begin{table}
\thispagestyle{empty}
\vspace*{-2.20truein}
\hspace*{-0.75truein}
\psfig{file=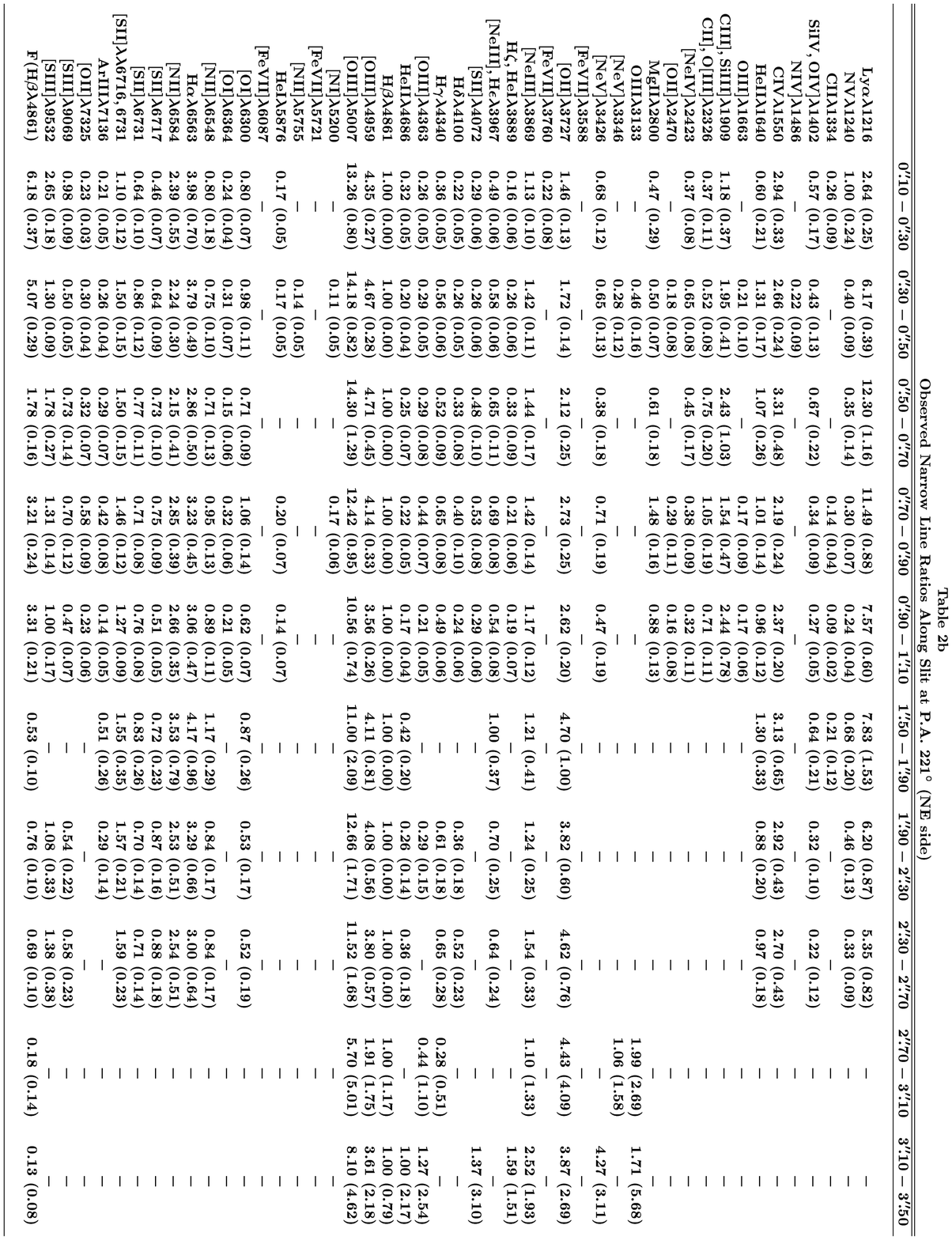}
\end{table}
\clearpage
\begin{table}
\vspace*{-2.0truein}
\hspace*{-1.0truein}
\psfig{file=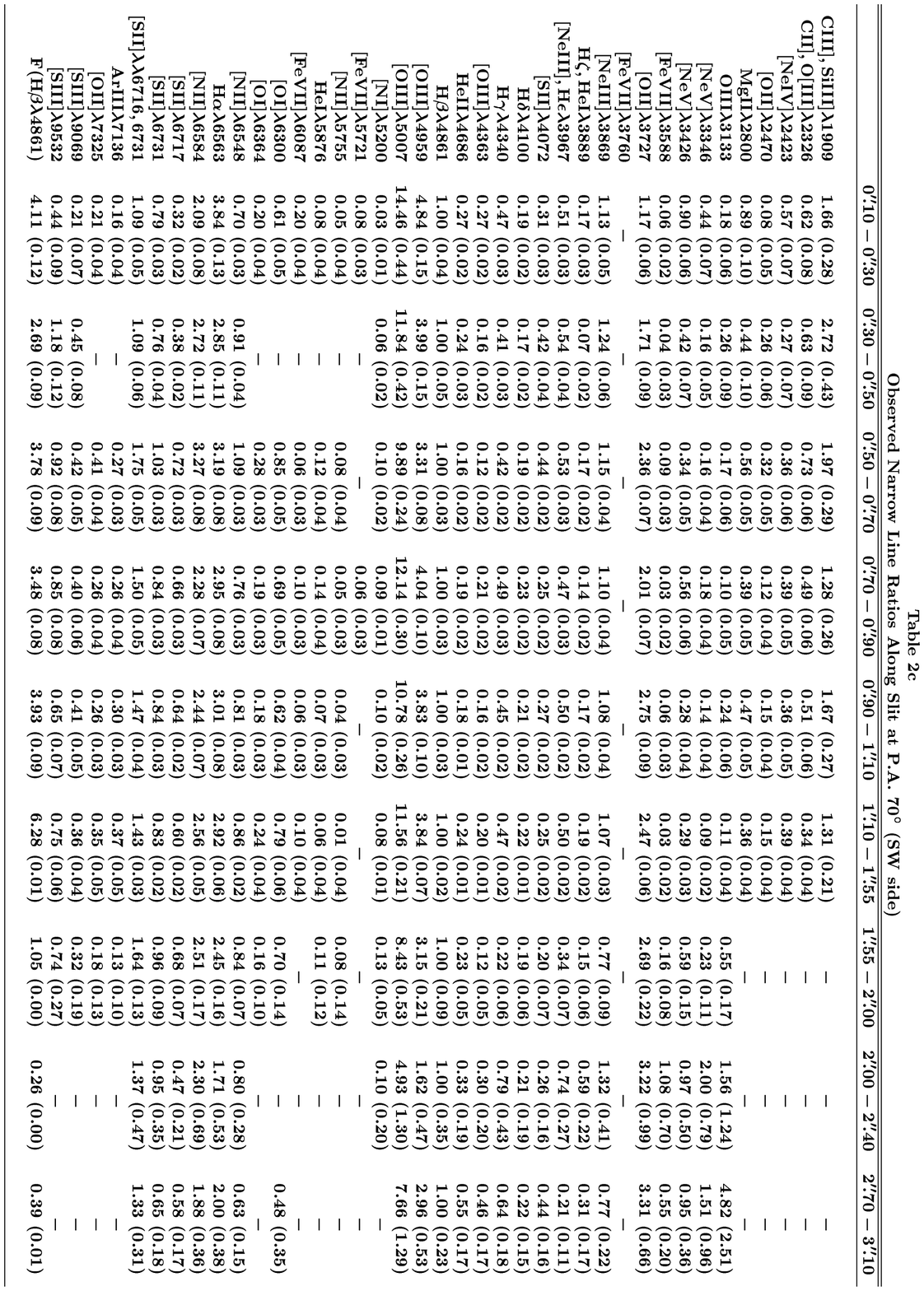}
\end{table}
\clearpage
\begin{table}
\vspace*{-2.0truein}
\hspace*{-1.0truein}
\psfig{file=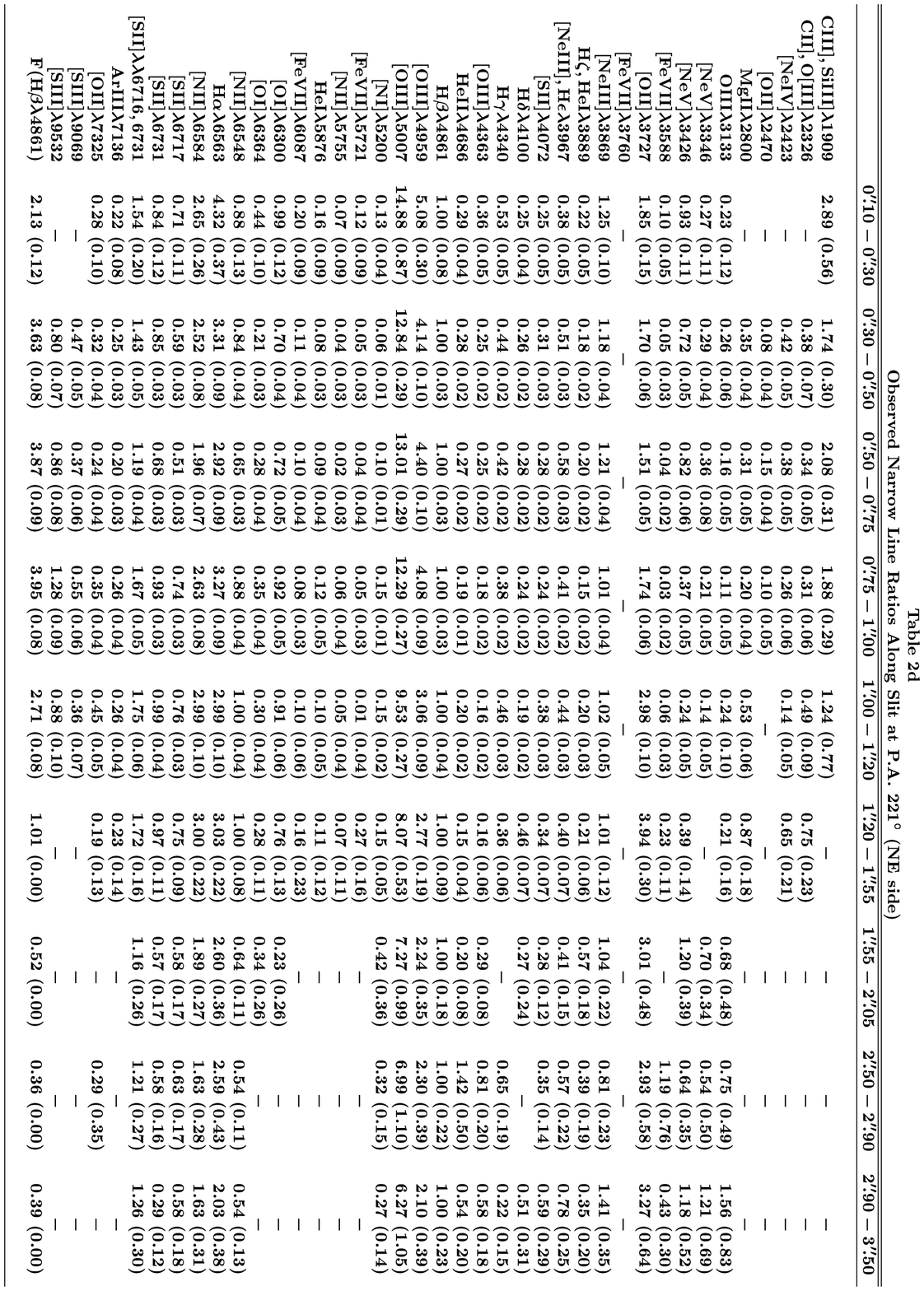}
\end{table}
\clearpage
\begin{table}
\vspace*{-2.0truein}
\hspace*{-0.75truein}
\psfig{file=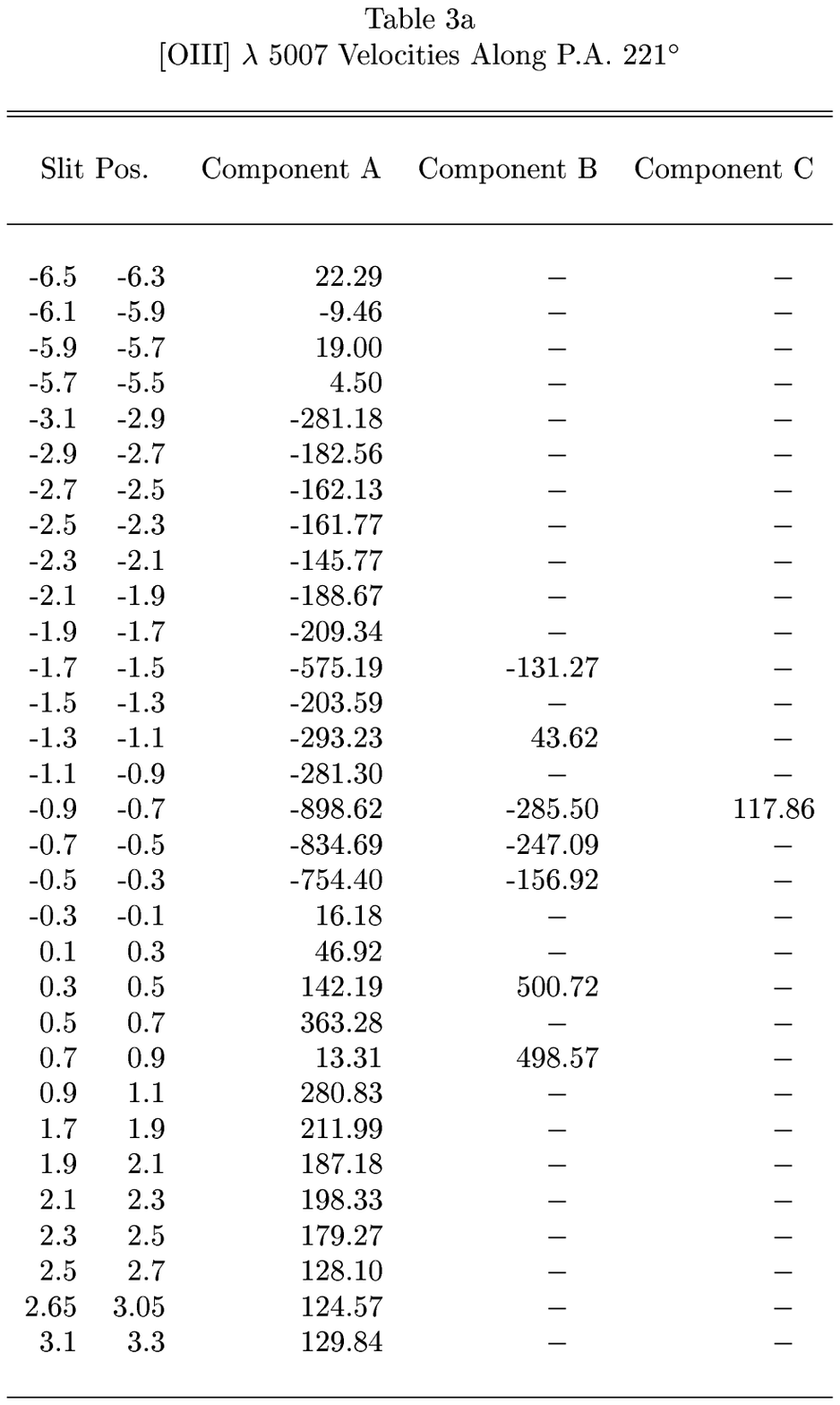}
\end{table}

\clearpage
\begin{table}
\vspace*{-2.0truein}
\hspace*{-0.75truein}
\psfig{file=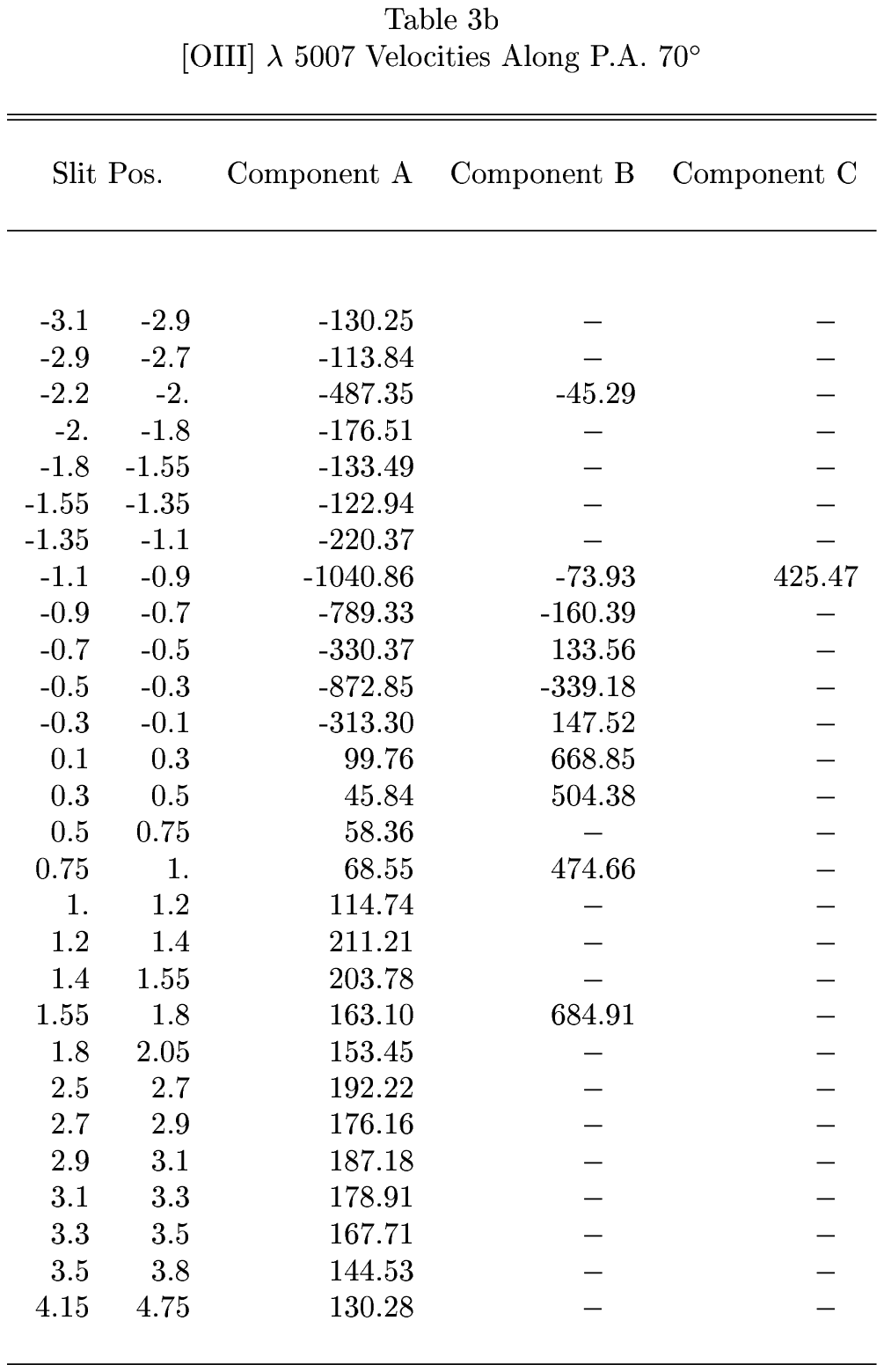}
\end{table}
\clearpage

\begin{table}
\vspace{-1.0truein}
\hspace{0.75truein}
\begin{tabular}{llcclcc}

\multicolumn{7}{c}{\bf TABLE 4a} \\
\multicolumn{7}{c}{\bf Cloud Identification P.A. 221\degx} \\
\hline\hline
& \multicolumn{3}{c}{SW Side} &\multicolumn{3}{c}{NE Side} \\  
& Extraction & Knot$^1$ & Region$^2$ & Extraction & Knot$^1$ &   
Region$^2$ \\
\hline
& $  0.1-0.3  $ & $-$ & $-$  & $  0.1-0.3  $ &  $-$ & $-$ \\
& $  0.3-0.5  $ & 26  &   N  & $  0.3-0.5  $ & 1 & $-$ \\
& $  0.5-0.7  $ & 23 & $-$ & $  0.5-0.7  $ & $-$ & $-$ \\
& $  0.7-0.9  $ & $-$ & $-$ & $  0.7-0.9  $ & 10   &  H    \\
& $  0.9-1.1  $ & 22 & $-$ & $  0.9-1.1  $ & 10 & $-$ \\
& $  1.1-1.3  $ & $-$ & $-$ & $  1.7-1.9  $ &  4 & $-$ \\
& $  1.3-1.5  $ & $-$ & $-$ & $  1.9-2.1  $ &  4  & $-$ \\
& $  1.5-1.7  $ & 25 & $-$ & $  2.1-2.3  $ &  4 & $-$ \\
& $  1.7-1.9  $ & 25 & $-$ & $  2.3-2.5  $ &  5 & $-$ \\
& $  1.9-2.1  $ & 25 & $-$ & $  2.5-2.7  $ &  5 & $-$ \\
& $  2.1-2.3  $ & 25 & $-$ & $  2.7-3.0  $ & $-$ & $-$ \\
& $  2.3-2.5  $ & $-$ & $-$ & $  3.1-3.3  $ & $-$ & $-$ \\
& $  2.5-2.7  $ & $-$ & $-$ \\
& $  2.7-2.9  $ & $-$ & $-$ \\
& $  2.9-3.1  $ & $-$ & $-$  \\
& $  5.5-5.7  $ & $-$ & $-$ \\
& $  5.7-5.9  $ & $-$ & $-$ \\
& $  5.9-6.1  $ & 33 & $-$ \\
& $  6.3-6.5  $ & 37 & $-$ \\
\hline
\multicolumn{7}{l}{$^1$Kaiser et al. 1999, $^2$ Hutchings et al. 1999}

\end{tabular}

\vspace{0.30truein}
\hspace{0.75truein}
\begin{tabular}{llcclcc}
\multicolumn{7}{c}{\bf TABLE 4b} \\
\multicolumn{7}{c}{\bf Cloud Identification P.A. 70\degx} \\
\hline\hline
& \multicolumn{3}{c}{SW Side} &\multicolumn{3}{c}{NE Side} \\  
& Extraction & Knot$^1$ & Region$^2$ & Extraction & Knot$^1$ &   
Region$^2$ \\
\hline
& $  0.1-0.3  $ & $-$ & $-$ & $  0.1-0.3  $ &  $-$ & $-$ \\
& $  0.3-0.5  $ & 26  &   N & $  0.3-0.5  $ &  11,6 & $-$ \\
& $  0.5-0.7  $ & 23 & $-$ & $  0.5-0.8  $ &  6 & $-$ \\
& $  0.7-0.9  $ & 19 & $-$ & $  0.8-1.0  $ &  9,12 & $-$ \\
& $  0.9-1.1  $ & $-$ &  D, D' & $  1.0-1.2  $ &  12 & $-$ \\
& $  1.1-1.4  $ & 16 & $-$ & $  1.2-1.4  $ &  $-$ & $-$ \\
& $  1.4-1.6  $ &  16 & $-$ & $  1.4-1.5  $ &  $-$ & $-$ \\
& $  1.6-1.8  $ &  $-$ & $-$ & $  1.5-1.8  $ &  3 & $-$ \\
& $  1.8-2.0  $ &  $-$ & $-$ & $  1.8-2.1  $ &  3 & $-$ \\
& $  2.0-2.2  $ &  $-$ & $-$ & $  2.5-2.7  $ &  14 & $-$ \\
& $  2.7-2.9  $ &  18 & $-$ & $  2.7-2.9  $ &  $-$ & $-$ \\
& $  2.9-3.1  $ &  18 & $-$ & $  2.9-3.1  $ &  $-$ & $-$ \\
&&&& $  3.1-3.3  $ &  $-$ & $-$ \\
&&&& $  3.3-3.5  $ &  $-$ & $-$ \\
&&&& $  3.5-3.8  $ &  $-$ & $-$ \\
&&&& $  4.2-4.7  $ &  29 & $-$ \\
\hline
\multicolumn{7}{l}{$^1$Kaiser et al. 1999, $^2$ Hutchings et al. 1999}

\end{tabular}

\smallskip

\end{table}

\clearpage

\clearpage

\clearpage

\clearpage

\clearpage

\clearpage

\clearpage

\clearpage

\clearpage

\clearpage

\clearpage

\clearpage

\begin{figure}
\vspace*{-1.5truein}
\hspace*{0.25truein}
\plottwo{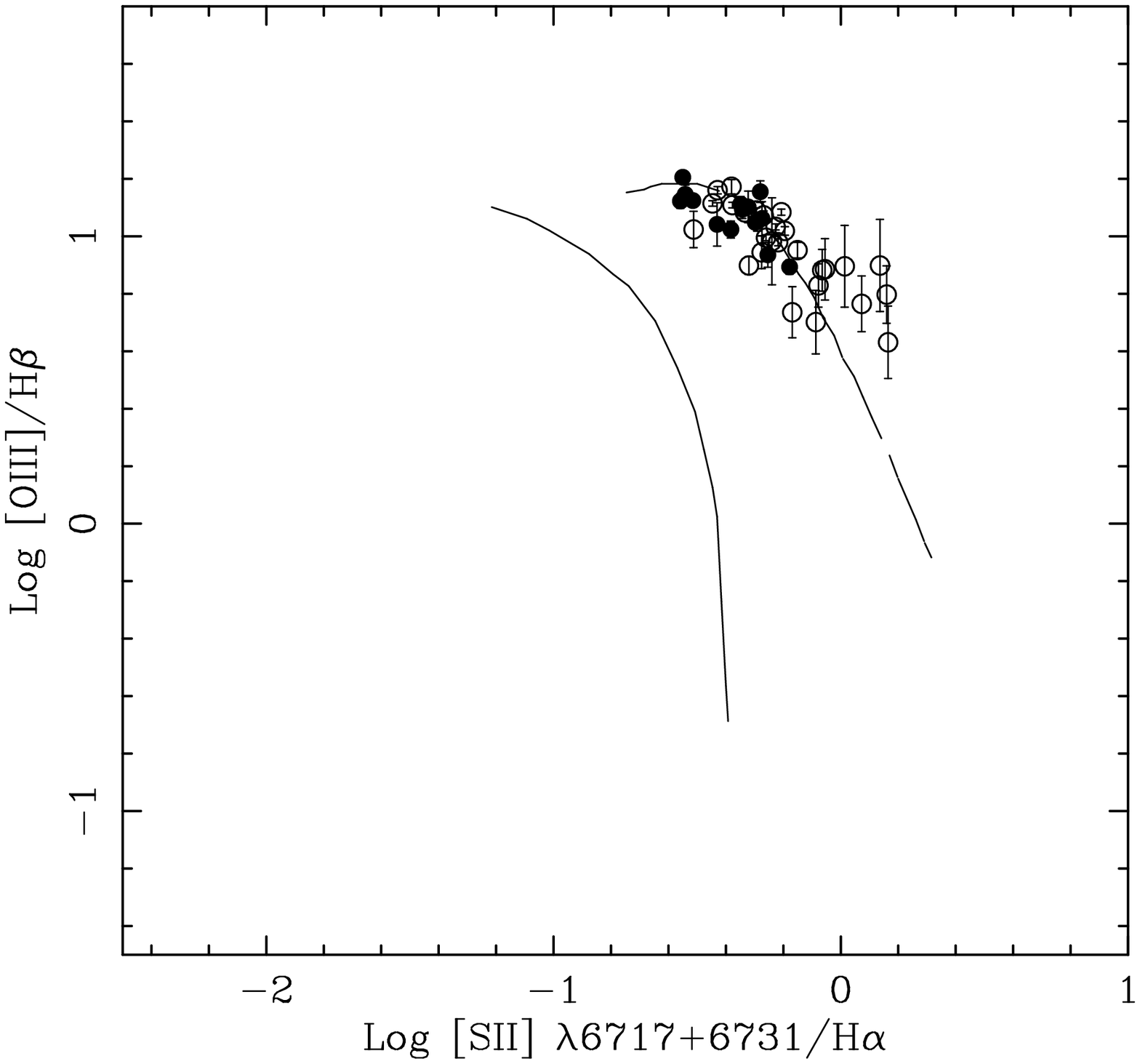}{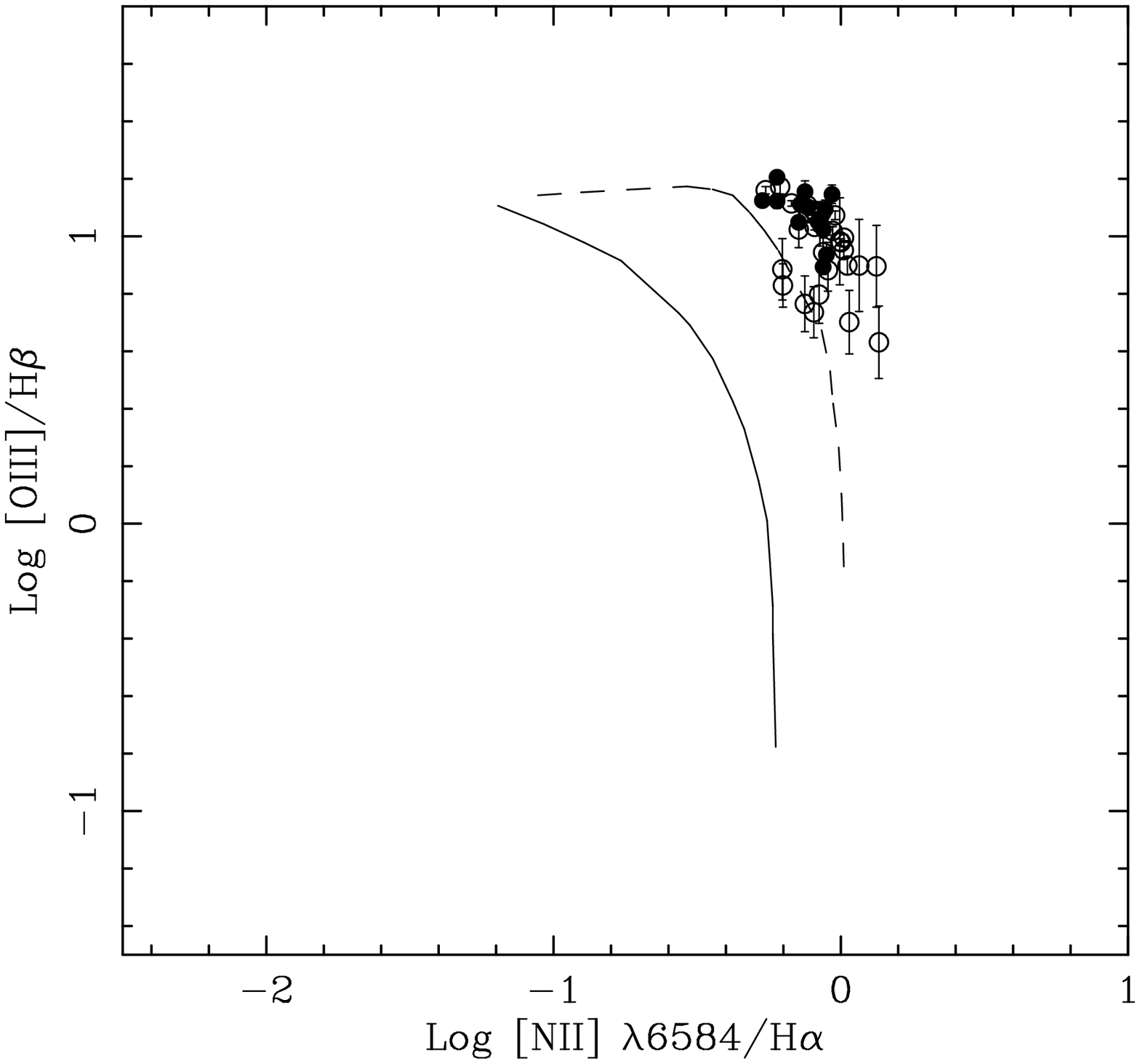}
\vspace{0.0truein}
\end{figure}

\begin{figure}
\vspace*{-1.5truein}
\hspace*{0.25truein}
\plotoney{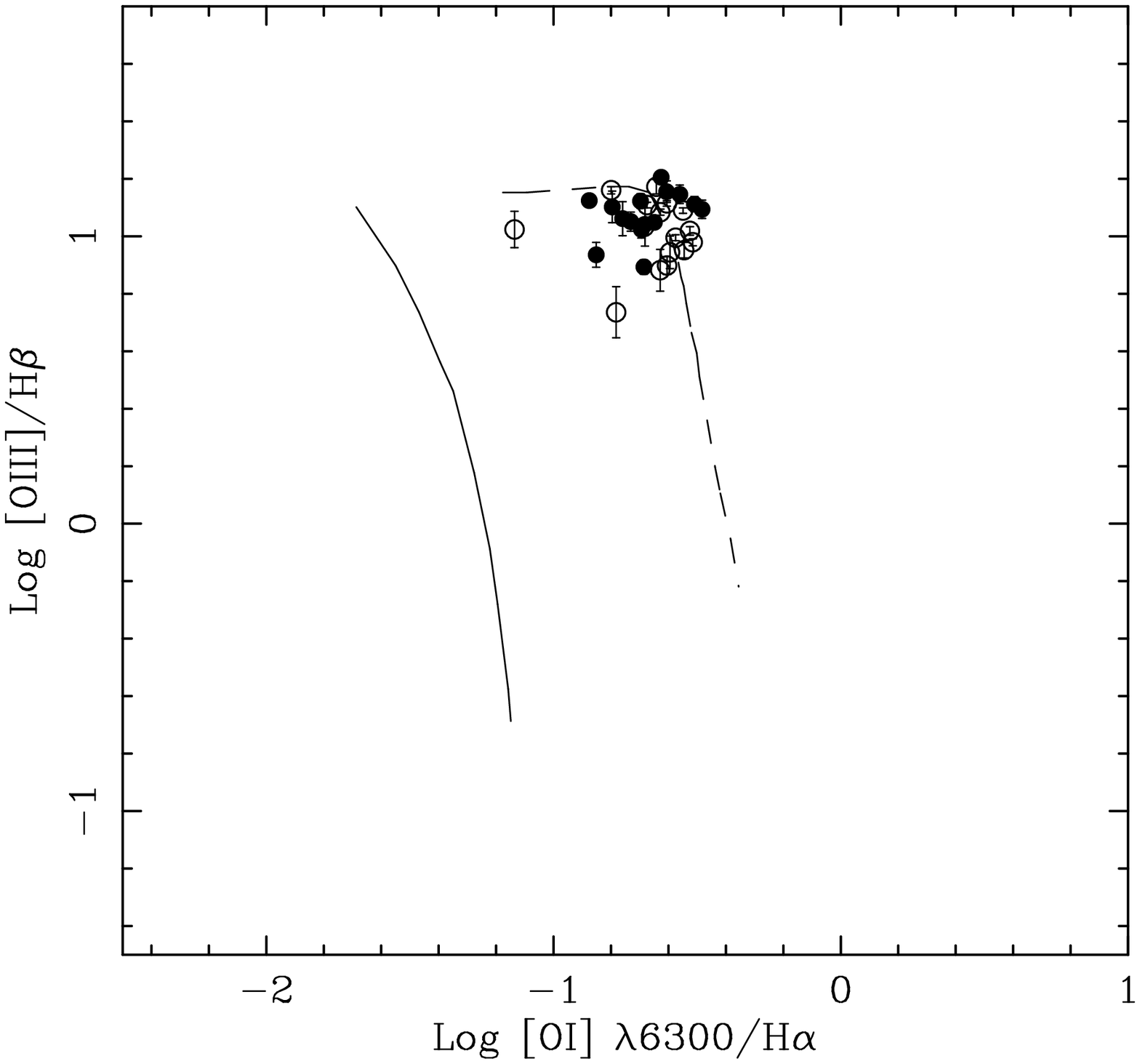}
\vspace{0.0truein}
\figcaption{Standard optical emission line ratio diagrams are plotted:
a, \Siia $+$ \Siib vs. \Oiii/\Hb; b, [NII] $\lambda 6584$/\Ha
vs. \Oiii/\Hb; c, [OI] $\lambda 6300$/\Ha vs. \Oiii/\Hb. In each
diagram the solid line shows the line separating star-forming regions
from AGN and the dashed line shows the photoionization models
of Ferland and Netzer (1983) for solar abundance. The results show
a compact distribution of points for the NGC 4151 NLR, suggesting
uniform excitation for the ensemble of clouds.
\label{ratioplots}}
\end{figure}
\clearpage

\begin{figure}
\vspace*{-1.5truein}
\hspace*{0.25truein}
\plottwo{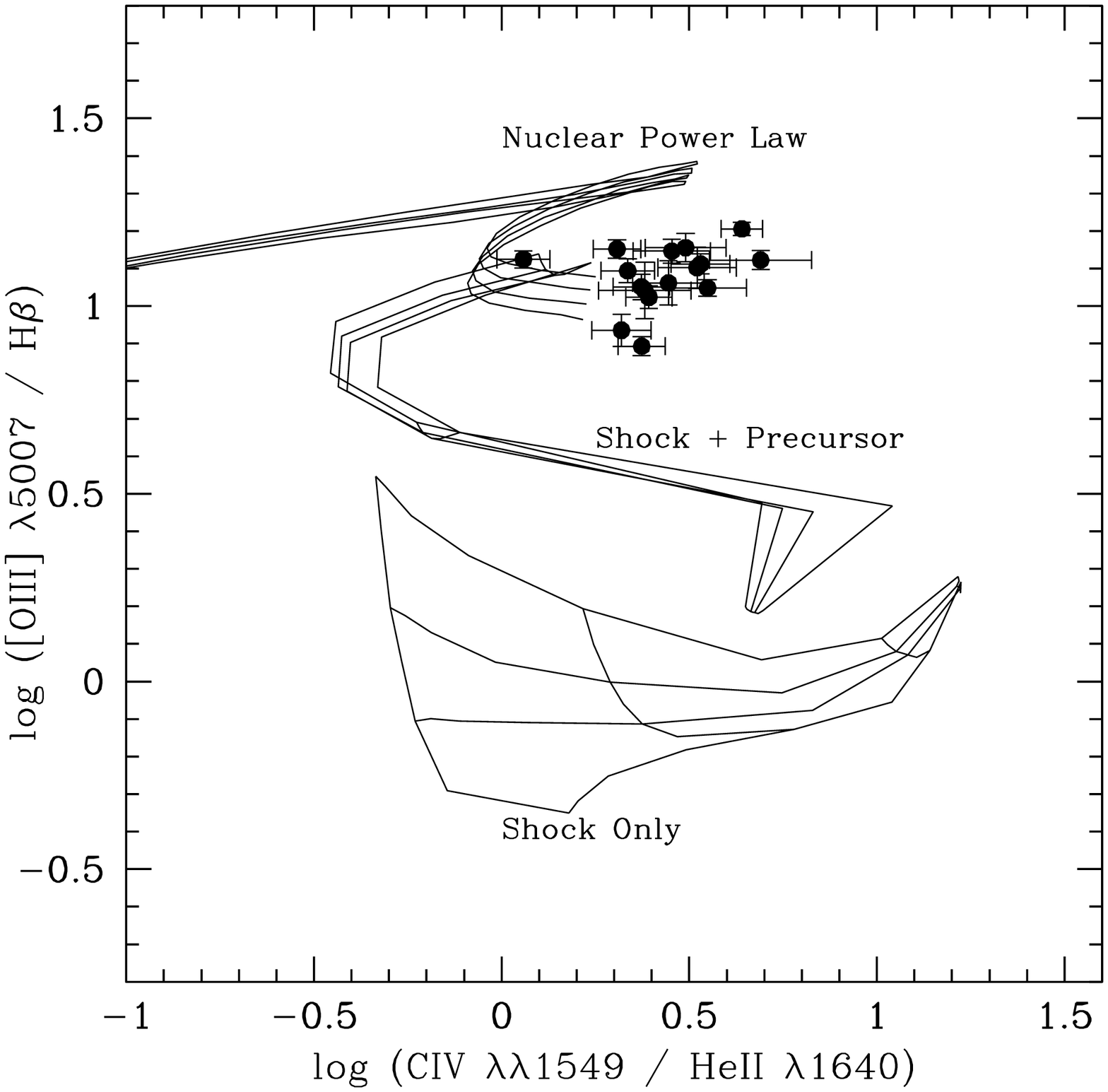}{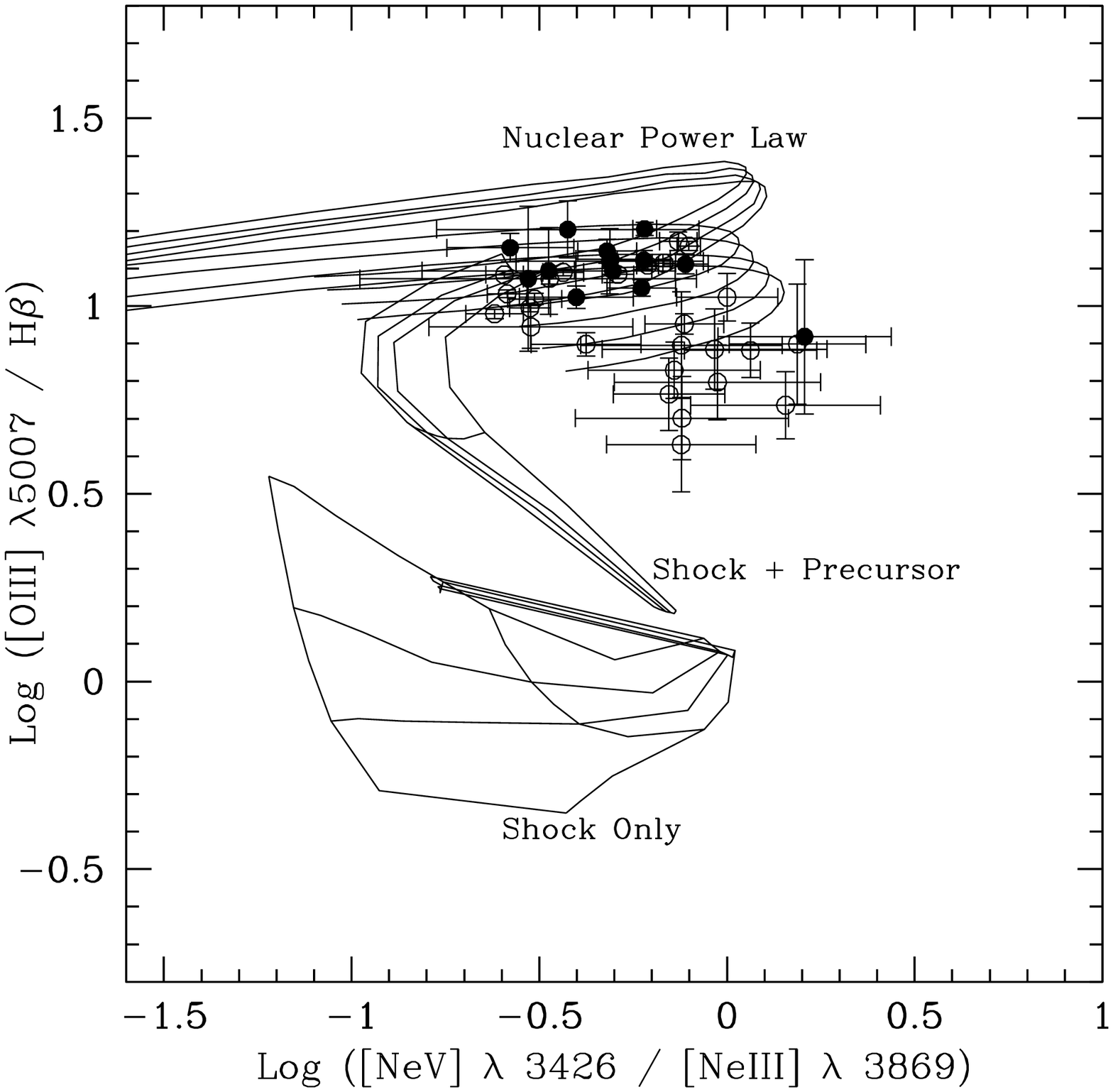}
\end{figure}

\begin{figure}
\vspace*{-1.5truein}
\hspace*{0.25truein}
\plotoney{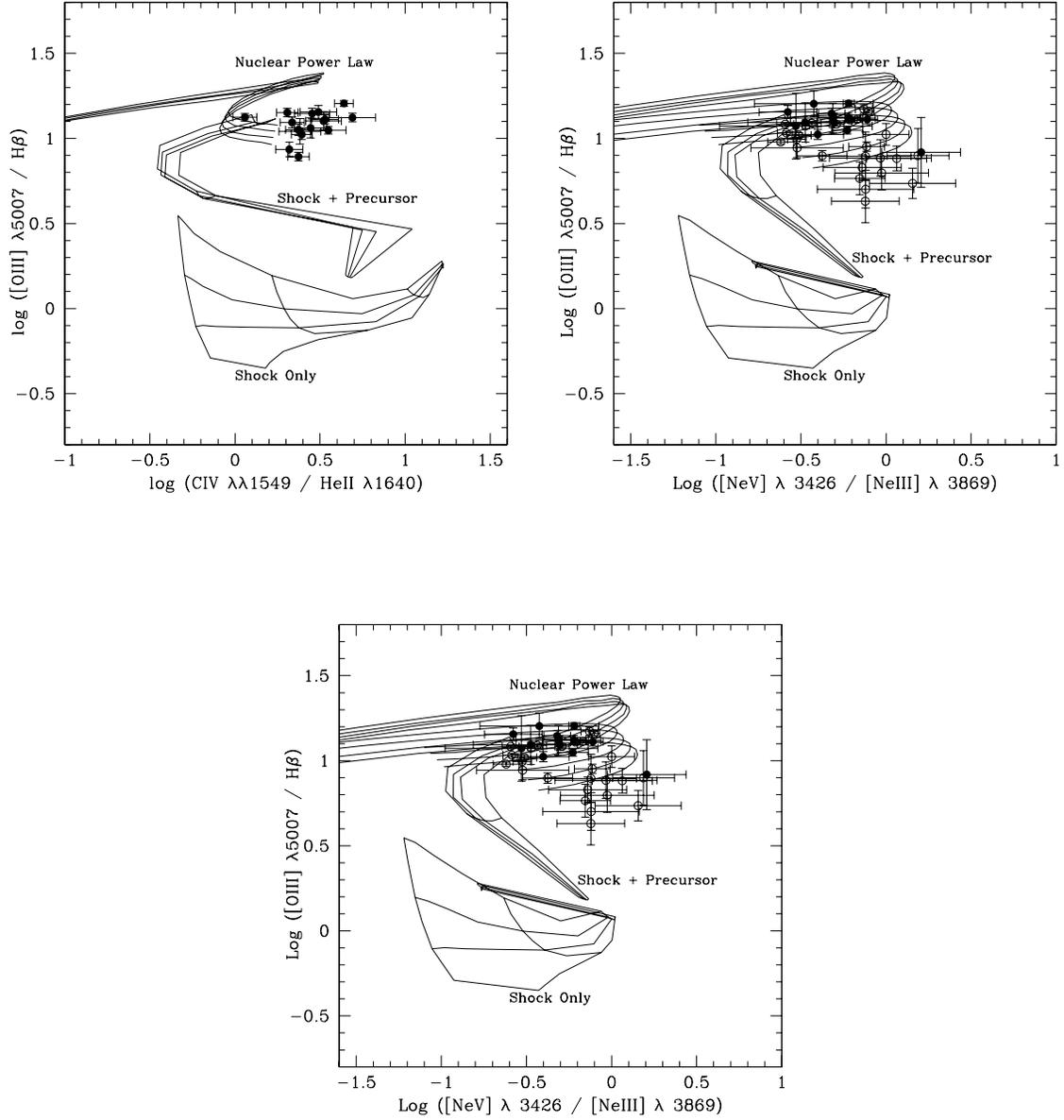}
\vspace{0.3truein}
\figcaption{Emission line ratio diagrams in the ultraviolet and near
ultraviolet including grids for shock ionization models, auto-ionizing
shock models and power-law ionization models calculated using the
MAPPINGS II code (Sutherland and Dopita, 1993, Allen {\it et al.}
1998). For the shock plus precursor models, the shock velocity
increases from 200 \kms to 500 \kms moving from low to high \Oiii/\Hb
ratios. The power-law models are for index $\alpha=-1$ and
$\alpha=-1.4$ ($f_{\nu} \sim \nu^{\alpha}$) and densities $n_e=100 \rm
cm^{-1}$ and $n_e=1000 \rm cm^{-1}$. The ionization parameter for
these models varies from roughly $U=10^{-2.5}$ to $U=1$ starting from
the left hand edge of each diagram. \label{shockgrids}}
\end{figure}

\newpage

\end{document}